\documentclass[
superscriptaddress,
twocolumn,
showpacs,preprintnumbers,
amsmath,amssymb,
aps,
prb,
]{revtex4-2}
\usepackage[colorlinks,
linkcolor=blue, urlcolor=blue, anchorcolor=blue, citecolor=blue]{hyperref}
\usepackage{graphicx}
\usepackage{dcolumn}
\usepackage{bm}
\usepackage{float}
\usepackage{multirow}
\usepackage{xcolor}
\usepackage{lipsum}  
\usepackage[normalem]{ulem}
\usepackage{titlesec}

\newcommand{\angstrom}{\textup{\AA}}
\begin{document}


\title{FeS$_2$ monolayer: a high valence and high-$T_{\rm C}$ Ising ferromagnet}


\author{Ke Yang}
\affiliation{College of Science, University of Shanghai for Science and Technology, Shanghai 200093, China}
 \affiliation{Laboratory for Computational Physical Sciences (MOE),
 State Key Laboratory of Surface Physics, and Department of Physics,
 Fudan University, Shanghai 200433, China}

 \author{Yaozhenghang Ma}
 \affiliation{Laboratory for Computational Physical Sciences (MOE),
 State Key Laboratory of Surface Physics, and Department of Physics,
 Fudan University, Shanghai 200433, China}

 \author{Lu Liu}
 \affiliation{Laboratory for Computational Physical Sciences (MOE),
 State Key Laboratory of Surface Physics, and Department of Physics,
 Fudan University, Shanghai 200433, China}

 \author{Yueyue Ning}
 \affiliation{College of Science, University of Shanghai for Science and Technology, Shanghai 200093, China}

\author{Di Lu}
 \affiliation{Laboratory for Computational Physical Sciences (MOE),
 State Key Laboratory of Surface Physics, and Department of Physics,
 Fudan University, Shanghai 200433, China}

\author{Yuxuan Zhou}
 \affiliation{Laboratory for Computational Physical Sciences (MOE),
 State Key Laboratory of Surface Physics, and Department of Physics,
 Fudan University, Shanghai 200433, China}

 \author{Zhongyao Li}
 \affiliation{College of Science, University of Shanghai for Science and Technology, Shanghai 200093, China}

\author{Hua Wu}
\email{Corresponding author. wuh@fudan.edu.cn}
\affiliation{Laboratory for Computational Physical Sciences (MOE),
 State Key Laboratory of Surface Physics, and Department of Physics,
 Fudan University, Shanghai 200433, China}
\affiliation{Shanghai Qi Zhi Institute, Shanghai 200232, China}
\affiliation{Collaborative Innovation Center of Advanced Microstructures,
 Nanjing 210093, China}

\date{\today}

\begin{abstract}

  Two-dimensional (2D) magnetic materials are of current great interest for their promising applications in spintronics. Strong magnetic coupling and anisotropy are both highly desirable for the achievement of a high temperature magnetic order. Here we propose the unusual high valent FeS$_2$ hexagonal monolayer as such a candidate for a strong Ising 2D ferromagnet (FM), by spin-orbital state analyses, first-principles calculations, and the renormalized spin-wave theory (RSWT). We find that very importantly, the high valent Fe$^{4+}$ ion is in the low-spin state ($t_{2g}^{4}$, $S$=1) with degenerate $t_{2g}$ orbitals rather than the high-spin state ($t_{2g}^{3}e_g^{1}$, $S$=2). It is the low-spin state that allows to carry a large perpendicular orbital moment and then produces a huge single ion anisotropy (SIA) of 25 meV/Fe. Moreover, the negative charge transfer character associated with the unusual high valence, strong Fe $3d$-S $3p$ hybridization, wide bands, and a small band gap all help to establish a strong superexchange. Indeed, our first-principles calculations confirm the strong FM superexchange and the huge perpendicular SIA, both of which are further enhanced by a compressive strain. Then, our RSWT calculations predict that the FM $T_{\rm C}$ is 261 K for the pristine FeS$_2$ monolayer and could be increased to 409 K under the compressive --5\% strain. The high $T_{\rm C}$ is also reproduced by our Monte Carlo (MC) simulations. Therefore, it is worth exploring the high-$T_{\rm C}$ Ising FMs in the high valent 2D magnetic materials with degenerate orbitals. 
  
\end{abstract}

\maketitle

\section*{I. Introduction}

Since the discovery of 2D FM CrI$_3$~\cite{Huang2017} and Cr$_2$Ge$_2$Te$_6$~\cite{Gong2017} in 2017, 2D magnetic materials have attracted great interest due to their promising applications in quantum devices and information technology~\cite{Li2019,Song2019,li2020,Burch_2018,song_2018}. 
According to the Mermin-Wagner theorem~\cite{MW}, magnetic anisotropy (MA) is essential for establishing a long-range magnetic order in 2D materials. 
Both the CrI$_{3}$ monolayer~\cite{Huang2017} and Cr$_{2}$Ge$_{2}$Te$_{6}$ bilayer~\cite{Gong2017} have a weak perpendicular MA, which was ascribed to an exchange anisotropy caused by the spin-orbit coupling (SOC) of the heavy ligand $p$ orbitals and their hybridization with the Cr 3$d$ orbitals~\cite{Lado2017, Xu2018, Kim_2019}. Here the octahedral Cr$^{3+}$ $S$=3/2 ion has a closed $t_{2g}^3$ shell, and its orbital singlet produces no single-ion anisotropy (SIA). In contrast, the VI$_3$ monolayer has an open V$^{3+}$ $t_{2g}^{2}$ shell, which allows for an unquenched orbital moment and a large SIA of 16 meV per V$^{3+}$ ion~\cite{Yang2020}. This accounts for the recent experimental observations of the large orbital moment and Ising magnetism in VI$_3$ monolayer~\cite{lin2021, Hao_2021, De2022}. As a large MA maintains a preferred magnetic orientation against the thermal fluctuation and excitation, it may stabilize a long-range magnetic order at a high temperature~\cite{Yang2021,Lu2022,Huang_2018_jacs,Xue_2019,liu2020, Huang2017,Gong2017,lin2021, Hao_2021, De2022}. To facilitate practical applications of the 2D magnetic materials, it is highly desirable to achieve both a strong magnetic coupling and anisotropy in them and thus their high ordering temperature.

\begin{figure}[H]
  \includegraphics[width=9cm]{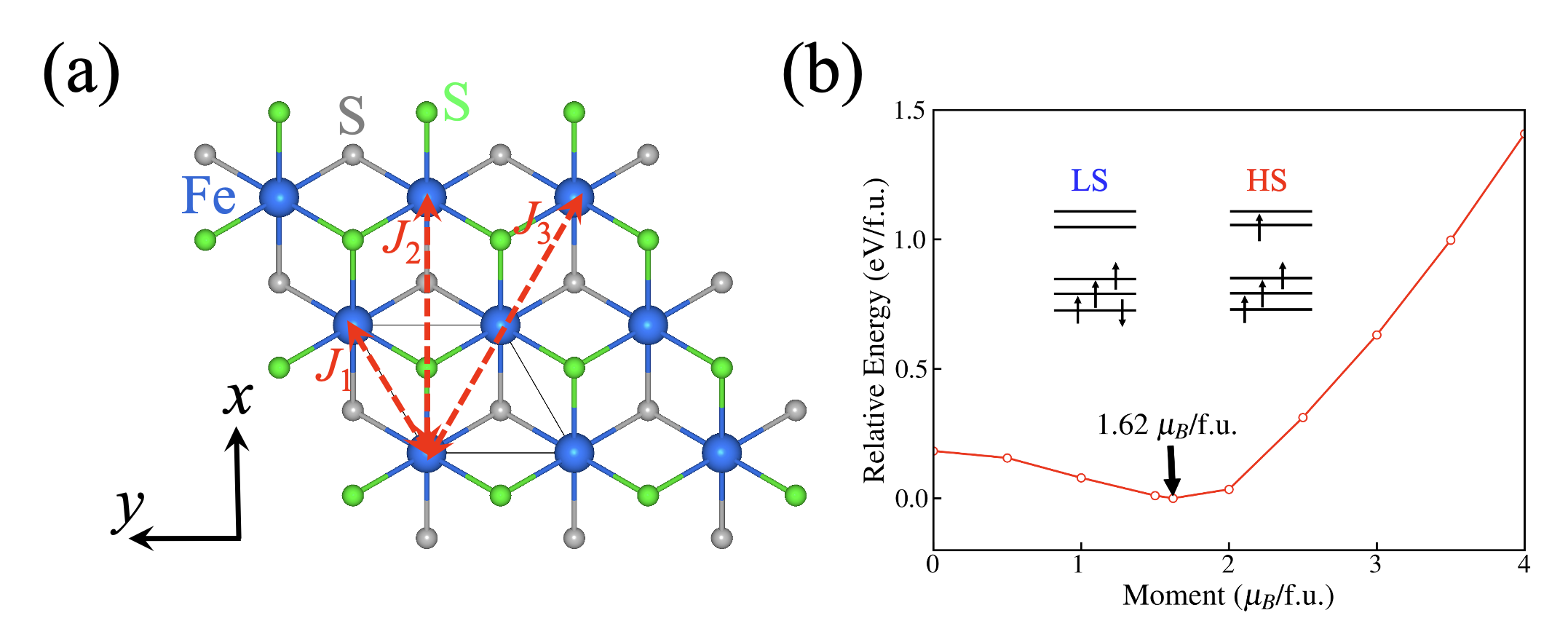}
  \centering
  \caption{(a) The crystal structure of FeS$_2$ monolayer. $J_1$, $J_2$, and $J_3$ refer to the first, second, and third nearest neighbor magnetic exchanges. (b) Fixed-spin-moment calculations imply the formal low-spin state with $S$=1 for Fe$^{4+}$.}
\label{FSM}
\end{figure}

Very recently, an unusual high valent FeS$_2$ monolayer in a triangular structure was successfully synthesized using a competitive-chemical-reaction-based growth mechanism~\cite{Zhou2022}. Its main structural units are the edge-sharing FeS$_6$ octahedra arranged into a delicate triangular network of Fe ions, as seen in Fig. \ref{FSM}(a). It was found to be an intriguing 2D FM semiconductor with a small band gap and a low Curie temperature ($T_{\rm C}\approx$ 15 K), and its analog with higher $T_{\rm C}$ may have potential applications in photodetectors and spintronic devices~\cite {Zhou2022}.

In contrast to a series of existing 2D magnetic materials having the normal valence states such as Cr$^{3+}$ in CrI$_3$~\cite{Huang2017} and Mn$^{2+}$ in MnBi$_2$Te$_4$~\cite{Otrokov_2019,Deng_2020}, FeS$_2$ monolayer is quite unique because of the unusual high valent Fe$^{4+}$ ions. The high valence transition metal (TM) compounds often have exotic electronic states, strong covalency between the TM and ligands, and surprising magnetic properties~\cite{Seki_2016,Hosaka_2015,goto_2021}. 
Although the Hund exchange favors a high-spin (HS) state with a maximal spin for $3d$ TM ions, such as Cr$^{3+}$ $S$=3/2, Mn$^{2+}$ $S$=5/2, Fe$^{2+}$ $S$=2, and Fe$^{3+}$ $S$=5/2, the octahedral Fe$^{4+}$ ion may well be in a low-spin (LS) $S$=1 state rather than the HS $S$=2 state, as the very strong hybridization between Fe$^{4+}$ $3d$ and S$^{2-}$ $3p$ orbitals could dominate over the Hund exchange and then stabilize the LS state. In addition, the strong hybridization and wide band effect would narrow the band gap of semiconductors or even produce a metallic behavior, and all this would favor the magnetic couplings, either superexchange via the virtual excitation across the small band gap or the itinerant magnetism in the metallic systems. Moreover, the formal LS Fe$^{4+}$ ($3d^4$, $S$=1) ion in the local FeS$_6$ octahedron has the $t_{2g}^{3\uparrow,1\downarrow}$ configuration, and the degenerate $t_{2g}$ triplet orbitals could carry the large orbital moment $L$=1 and produce a strong SIA. Then, the pristine FeS$_2$ monolayer could have both the strong magnetic coupling and anisotropy, and consequently it could possess a high ordering temperature suitable for promising spintronic applications. Furthermore, we rationalize the recent experimental low $T_{\rm C}$ by demonstrating the S vacancy effect as seen below, and thus propose that the native high $T_{\rm C}$ of FeS$_2$ monolayer may be restored by removing the S vacancies.

The above pictures motivate us to study the electronic structure and magnetic property of FeS$_2$ monolayer, using the first-principles calculations, crystal-field level and spin-state diagrams, charge-transfer type superexchange analysis, and RSWT calculations. Indeed, our work confirms that the unusual high valent Fe$^{4+}$ ion is in the LS $S$=1 state and possesses a large out SIA-of-plane orbital moment of about 1 $\mu_{\rm B}$ and thus a huge perpendicular of 25 meV/Fe. Moreover, we find that FeS$_2$ monolayer is a charge-transfer type semiconductor with a tiny band gap and has considerably strong FM couplings. Thus, the pristine FeS$_2$ monolayer would have quite high $T_{\rm C}$ which is even increased above room temperature under a few percent compressive strain according to our RSWT calculations and MC simulations. Therefore, this prediction calls for an experimental study, and the high valent van der Waals magnetic materials with degenerate orbitals are worth exploring in the search of the 2D high-$T_{\rm C}$ Ising ferromagnets.

\section*{II. Computational Details}

We perform density functional theory (DFT) calculations using the Vienna \textit{ab} initio simulation package (VASP)~\cite{VASP}.
The kinetic energy cutoff for plane wave expansion is set to 500 eV. The Monkhorst-Pack grid of 11$\times$11$\times$1 is used for 1$\times$1 planar unit cell, and (3$\times$3$\times$1) k-mesh for 4$\times$4 planar supercell. The experimental lattice constants of $a$ = $b$ = 3.23 $\angstrom$~\cite{Zhou2022} for FeS$_2$ monolayer are used in our calculations, as the DFT optimized lattice constants $a$ = $b$ = 3.20 $\angstrom$ are almost the same as them.
The atomic positions are fully relaxed till the force on each atom is less than 0.01 eV/\AA~and the total energy minimization is performed with a tolerance of 10$^{-5}$ eV.
A vacuum space larger than 15 \AA~is employed to avoid periodic image interactions between FeS$_2$ layers.
To account for the electron correlation of Fe 3$d$ states, the local-spin-density approximation plus Hubbard $U$ (LSDA + $U$) calculations are performed using the common value of Hubbard $U$ = 5 eV and Hund exchange $J_{\rm H}$ = 1 eV, and moreover, the hybrid functional calculations are carried out. Both the calculations give very similar band structures as seen below and also reproduce the experimental small gap. Furthermore, we test the $U$ values in the reasonable range of 4-6 eV and find that our prediction of the quite high $T_{\rm C}$ Ising FM remains unchanged as seen below.

To check the ground state and possible metastable states of the spin-orbital system, we use the open-source software developed by Watson~\cite{Allen_2014} to construct the occupation number matrices for multiple spin-orbital states, see more details in Section I of Supplemental Material (SM)~\cite{SM} (see also references ~\cite{Zhou2022,WIEN2k,Becke_1993,Becke_1996, Fock25, hf_1,hf_2,li_2018,Metropolis} therein). Then, the LSDA+$U$ calculations read those occupation number matrices and yield the orbitally dependent/polarized potential in each iteration of electronic steps. Furthermore, those occupation number matrices are updated and read iteratively in self consistent calculations till a full electronic relaxation. Normally, in such calculations the ground state and several low-energy metastable states can be stabilized as they are, but some high-energy metastable states are too unstable to converge to the ground state or a low-energy metastable state. Then, one can determine the ground state of spin-orbital system in a reliable way using such calculations, as demonstrated in many previous works~\cite{Yang2020, Lu2022, liu2020, Yao_2021, Ou_2015}. For example, such calculations can well reproduce the experimental crystal field level splitting and orbital excitation energy, and in this work their accuracy is proven by reproducing the SOC strength of the Fe$^{4+}$ 3$d$ electrons as seen below.
Note also that the SOC is included for Fe and S atoms by the second-variational method with scalar relativistic wave functions.
The magnetic phase transition of the FeS$_2$ monolayer is probed using RSWT and MC simulations, see more computational details in SM~\cite{SM}.

\section*{III. Results and Discussion}

\begin{figure}[t]
  \includegraphics[width=8cm]{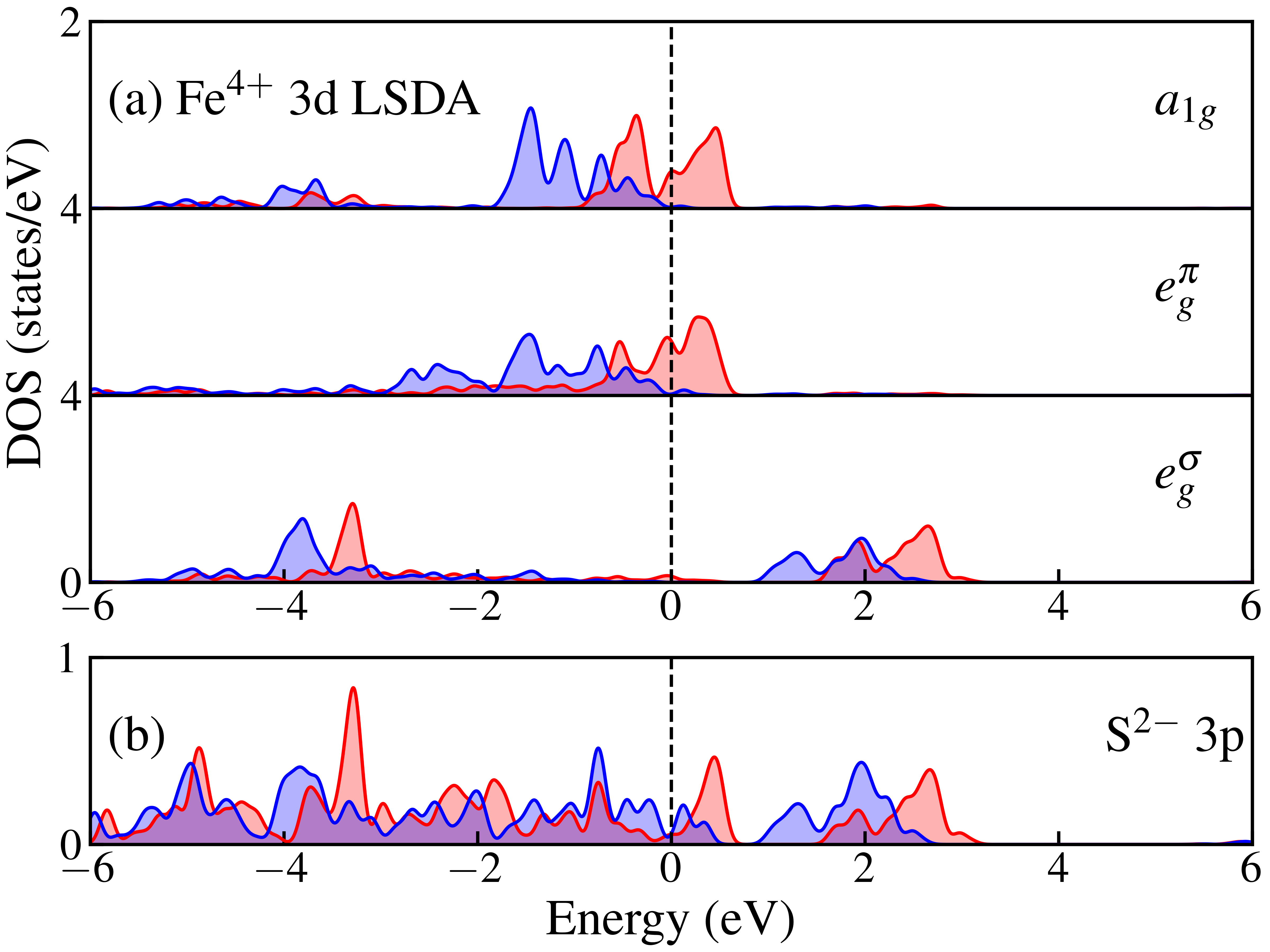}
  \centering
  \caption{(a) Fe 3$d$ and (b) S 3$p$ DOS by LSDA. The blue (red) curves stand for the up (down) spin channel. The Fermi level is set at zero energy.
  }
\label{LSDA_DOS}
\end{figure}

\subsection*{A. The Low-Spin State of The Fe$^{4+}$ Ion}

To clearly see the crystal field effect, exchange splitting, electron correlation, and the crucial SOC effects, we present and discuss below the LSDA and LSDA+SOC+$U$ calculations.
We first carry out LSDA calculations to see the basic electronic structure of FeS$_2$ monolayer. The Fe $3d$ states in the local FeS$_6$ octahedron split into the higher-lying $e_g^{\sigma}$ doublet and the lower $t_{2g}$ triplet with the energy separation of about 2 eV, and the $t_{2g}$ further splits into $a_{1g}$ singlet and $e_g^{\pi}$ doublet in the global trigonal crystal field, as seen in Fig.~\ref{LSDA_DOS} and in section I of SM~\cite{SM}. The $a_{1g}$ and $e_g^{\pi}$ are almost degenerate and spin-polarized, with the up-spin channel being fully filled but the down-spin one partially occupied, and the $e_g^{\sigma}$ doublet is formally unoccupied, with the antibonding state lying around 2 eV above the Fermi level. Obviously, there is a strong hybridization between the Fe $3d$ and S $3p$ states, making their broad bands distributing over a large energy range from --6 eV to 3 eV. It is the strong hybridization that produces a huge $pd\sigma$ bonding-antibonding splitting of about 6 eV (about --4 eV vs 2 eV) for the Fe $3d$ $e_g^{\sigma}$ states. There are a lot of holes on the S $3p$ states, featuring the negative charge-transfer behavior in such unusually high valent compounds~\cite{Mizokawa1991,Bisogni2016}. The calculated local spin moment is 1.50 $\mu_{\rm B}$ per Fe$^{4+}$, implying the low-spin $S$=1 state with the formal $t_{2g}^{3\uparrow,1\downarrow}$ configuration. Owing to the strong Fe $3d$-S $3p$ hybridization, the S $3p$ fat orbitals/delocalized states get also somewhat spin-polarized, having 0.03 $\mu_{\rm B}$ per S within its muffin-tin sphere and 0.06 $\mu_{\rm B}$ in the interstitial region per formula unit (fu). 
All those spin moments add up to the total spin moment of 1.62 $\mu_{\rm B}$/fu, but this moment is still reduced from the formal Fe$^{4+}$ $S$=1 state, just by the electron itineracy in the LSDA metallic band structure originating from the strong Fe $3d$-S $3p$ hybridization.

To double-check the possible LS ground state of the formal Fe$^{4+}$ ion, we perform fixed-spin-moment calculations and compute the LSDA total energies for the total spin moments ranging from the nonmagnetic $S$=0 state via the LS $S$=1 to the HS $S$=2 state. As seen in Fig. \ref{FSM}(b), the obtained results show that the ground state indeed has a total spin moment close to 2 $\mu_{\rm B}$/fu (i.e., the LS $S$=1 state), and that the HS $S$=2 state is extremely unstable, lying above the LS state too much by about 1.5 eV/fu. All the above results show that the unusual high valent Fe$^{4+}$ ion is in the formal LS $S$=1 state rather than the HS $S$=2 state.

\begin{figure}[t]
  \centering
\includegraphics[width=9cm]{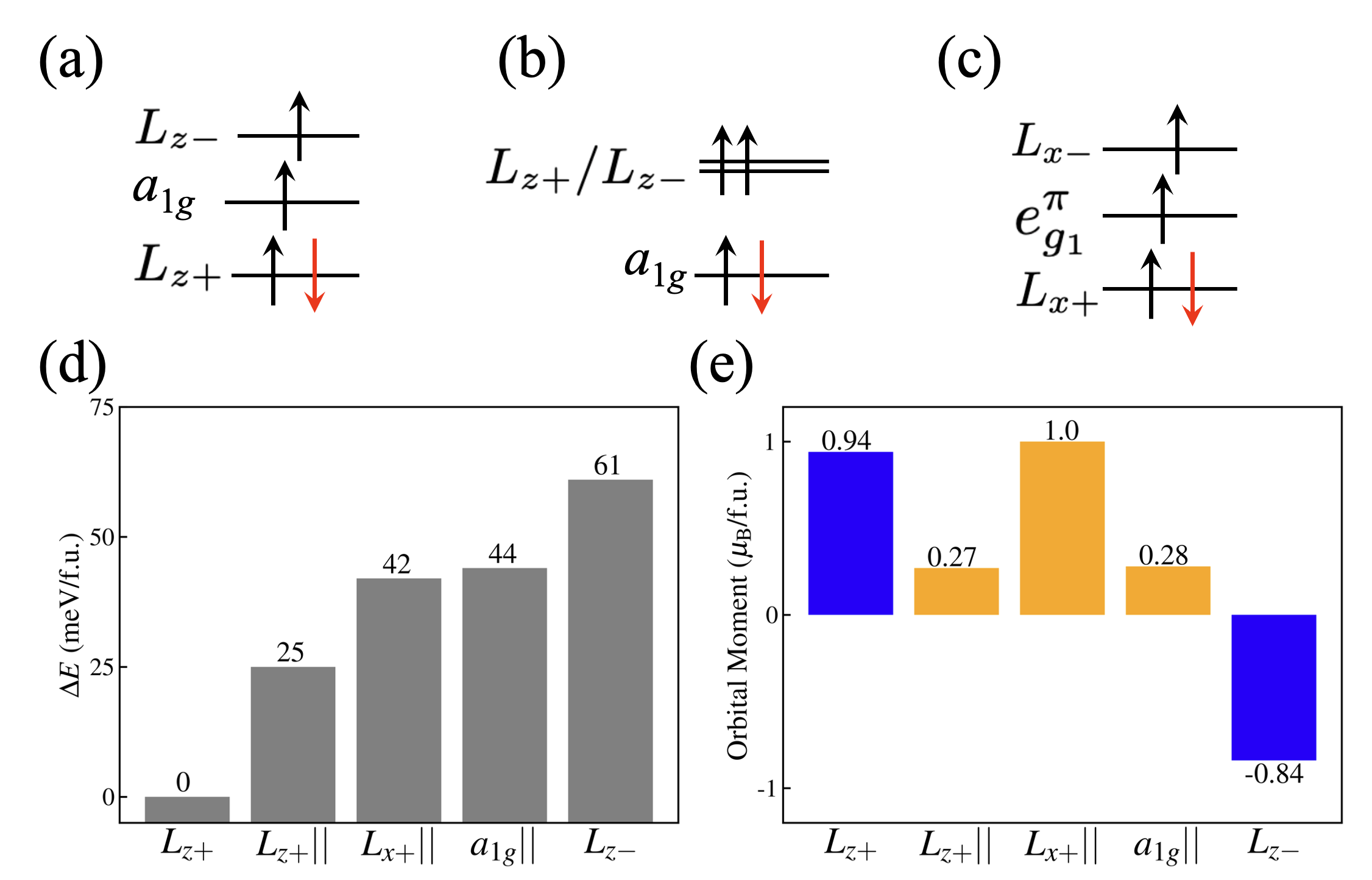}
\centering
 \caption{Crystal field level diagrams for the LS Fe$^{4+}$ $S$ = 1 ion in different configuration states: the single down-spin electron occupies the (a) $L_{z+}$, (b) $a_{1g}$, and (c) $L_{x+}$ states. (d) the relative total energies $\Delta$\textit{E} (meV/fu), and (e) orbital moments (blue color for out-of-plane and yellow for in-plane) for the FeS$_2$ monolayer in different states by LSDA+SOC+$U$. The symbol $\parallel$ in the state labellings marks the in-plane magnetization, in comparison with other states with out-of-plane magnetization.
}
\label{states}
\end{figure}

\subsection*{B. The $L_z$=1 Ground State and Huge SIA}

The LS Fe$^{4+}$ ion has the formal $t_{2g}^{3\uparrow,1\downarrow}$ configuration, and the nearly degenerate $a_{1g}$ and $e_{g}^{\pi}$ orbitals in the global trigonal crystal field could make the Fe$^{4+}$ ion stay in different orbital states, as seen in Fig. \ref{states}. Actually, all these states have the same full filling of the up-spin $t_{2g}$ orbitals, but differ only in the occupation of the down-spin $t_{2g}$ which is therefore used for labeling of those states. 
The different combinations of the $a_{1g}$ singlet and $e_{g}^{\pi}$ doublet in case of the corresponding orbital degeneracy, yield the $L_{z\pm}$, $L_{x\pm}$, or $L_{y\pm}$ state with an orbital moment of $\pm$1 $\mu_{\rm B}$ along the $z$, $x$, or $y$-axis, respectively, see Section I in SM for more details~\cite{SM}. Note that the $a_{1g}$ orbital singlet state in Fig. \ref{states}(b) formally has no orbital moment but could carry a small one due to the SOC mixing of the $a_{1g}$ and $e_{g}^{\pi}$. Moreover, owing to the actual trigonal crystal field splitting between the $a_{1g}$ singlet and $e_{g}^{\pi}$ doublet, the real $L_{z\pm}$ and $L_{x\pm}$ states obtained below could have the orbital moment different from $\pm$1 $\mu_{\rm B}$ which is also spin-orientation dependent due to the SOC effect, see Fig.~3(e). By inclusion of the SOC and electron correlation effects of Fe $3d$ orbitals, we carry out LSDA+SOC+$U$ calculations to determine which one in Fig. \ref{states}(d) is the spin-orbital ground state of FeS$_2$ monolayer. Indeed, all these states can be stabilized in our calculations guided by the initialized occupation number matrices and by the subsequent full electronic relaxation.

\begin{figure}[t]
  \includegraphics[width=8cm]{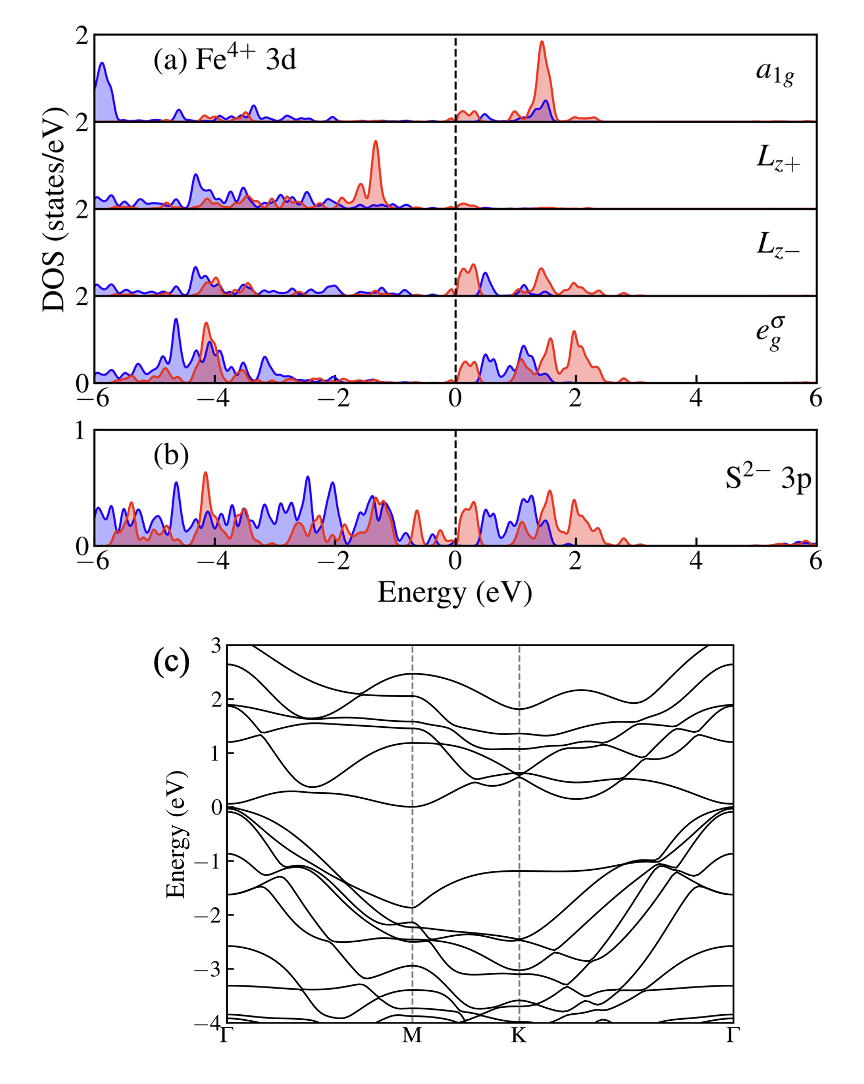}
   \caption{(a) Fe 3$d$ and (b) S 3$p$ DOS of FeS$_2$ monolayer in the $L_{z+}$ ground state by LSDA+SOC+$U$. The blue (red) curves stand for the up (down) spin channel. The Fermi level is set at zero energy. (c) The corresponding band structure with a tiny band gap of 40 meV.
  }
  \label{LSDA_U_SOC_DOS}
  \end{figure}

All the solutions are semiconducting with a tiny band gap, and the corresponding results are displayed in Fig. \ref{LSDA_U_SOC_DOS} and Section II in SM~\cite{SM}. 
The ground state is the $L_{z+}$. It has the total spin moment of 1.98 $\mu_{\rm B}$ and the parallel orbital moment of 0.94 $\mu_{\rm B}$ both along the $z$-axis (Figs. \ref{states}a and \ref{LSDA_U_SOC_DOS}). The $L_{z-}$ state, where the orbital moment is antiparallel to the total spin moment, would lose the SOC energy and the increasing total energy of 61 meV/fu [$\Delta E$=$\zeta$($\Delta l_z$)$s_z$=$\zeta\cdot 2\cdot 1/2$] leads to the estimate of the SOC parameter $\zeta$ = 61 meV for the formal Fe$^{4+}$ ion, see Fig. \ref{states}(d). As the $L_{z+}$ orbital moment firmly fixes via the SOC the parallel spin moment along the $z$-axis, and if the spin moment could be flipped into the $xy$ plane, the total energy would rise up by 25 meV/fu (Fig. \ref{states}(d)), showing a huge SIA energy of 25 meV/Fe which approaches the ideal $\zeta$/2. Moreover, we obtain the $L_{x+}$ state and it has both the $x$-axis spin moment of 2 $\mu_{\rm B}$ and the orbital moment of 1 $\mu_{\rm B}$, but its total energy is higher than the $L_{z+}$ ground state by 42 meV/fu, see Figs. \ref{states}(d) and (e). We also get the $a_{1g}$ state and it has a finite in-plane orbital moment of 0.28 $\mu_{\rm B}$ and lies above the $L_{z+}$ ground state by 44 meV/fu. 

Then, in FeS$_2$ monolayer, the formal Fe$^{4+}$ ion is in the low-spin $t_{2g}^4$ state with $S$ = 1 and $L_z$ = 1, and those occupied states are separated from other unoccupied Fe 3$d$ states by electronic Coulomb correlations, which essentially determine the insulating behavior of FeS$_2$ monolayer, see Fig. 4(a). The negative charge-transfer character and strong Fe 3$d$-S 3$p$ hybridization, both associated with the unusually high Fe$^{4+}$ valence state, give rise to broad bands which eventually reduce the above gap drastically to a minor semiconducting gap as seen in Figs. 4(b) and 4(c).

Indeed, the energy splitting between the $L_{z+}$ and $L_{z-}$ orbitals arises from the SOC effect. In addition to this, electron correlation plays a significant role in determining the energy splitting between the occupied and unoccupied Fe 3$d$ states, as shown in Fig. 4(a). In this work, we stabilize the spin-orbital ground state and several metastable states using LSDA+SOC+$U$ calculations, where the ‘large’ energy splitting between the occupied and unoccupied Fe 3$d$ states are always present due to the Hubbard $U$. In order to determine the spin-orbital excitation energy of the Fe 3$d$ states, e.g., the $L_{z+}$/$L_{z-}$ orbital splitting by SOC, we cannot use the DOS results but instead use the computed total energy differences as shown in Fig. 3(d). Our results show that the SOC $L_{z+}$/$L_{z-}$ splitting is 61 meV, which just reflects the SOC strength $\zeta$ of the Fe 3$d$ state. Actually, the $\zeta$ parameter of ionic Fe 3$d$ state is known to be about 60-70 meV, and the present agreement reflects the good accuracy of our calculations.

After investigating all the spin-orbital states of the unusual high-valent and the LS Fe$^{4+}$ ions, we find that the ground state is $L_{z+}$, see Fig. \ref{states}(a).
It has the nominal large orbital moment of 1 $\mu_{\rm B}$ which is parallel to the total spin moment of 2 $\mu_{\rm B}$, and has a small insulating gap of 40 meV as shown in Fig. \ref{LSDA_U_SOC_DOS}(c), which is in agreement with the observed small gap insulating behavior of FeS$_2$ monolayer~\cite{Zhou2022}. 
Note that this small gap band structure (Fig. 4(c)) is also well reproduced by a hybrid functional calculation which yields a very similar band structure, as seen in Fig. S6 of SM~\cite{SM}.
Moreover, the $L_{z+}$ ground state has a huge SIA energy of 25 meV/Fe and strongly favors the perpendicular magnetization. Therefore, FeS$_2$ monolayer could be an emerging 2D Ising magnet.

\subsection*{C. FM Couplings and High $T_{\rm C}$}

We now study the magnetic properties of FeS$_2$ monolayer. Besides the above calculated FM state, we also calculate three different AF states depicted in Fig. S7 of SM~\cite{SM}. All the calculations are based on the LS $L_{z+}$ spin-orbital ground state.
The results show that the FM solution is the ground state and the three exchange parameters $J_1$ = 5.81 meV, $J_2$ = 2.01 meV, and $J_3$ = 1.05 meV are all FM, see section IV in SM for more details~\cite{SM}.
For 2D FM semiconductors and insulators, normally a superexchange plays a dominant role in establishing the magnetic couplings, and the hybridization between the magnetic transition-metal ions and the ligand ions has a strong impact on the strength of the superexchange couplings. The largest $J_1$=5.81 meV can be qualitatively explained by the charge-transfer type superexchange, and the large distance $J_3$ of 1.05 meV is still sizeable due to the strong Fe $3d$-S $3p$ hybridization and the tiny band gap (in this unusually high valent system) both of which facilitate the long-range FM coupling. As FeS$_2$ monolayer is a tiny-gap charge-transfer semiconductor, the virtual excitations from S $3p$ to Fe $3d$ associated with the superexchange are energetically cheap. Moreover, the strong Fe $3d$-S $3p$ hybridizations yield large hopping parameters. Then the two major superexchange channels, counting the large $pd\sigma$ and medium $pd\pi$ hybridizations, contribute to the FM $J_1$ coupling, for more details see Section V in SM~\cite{SM}.

\begin{figure}[t]
  \includegraphics[width=8cm]{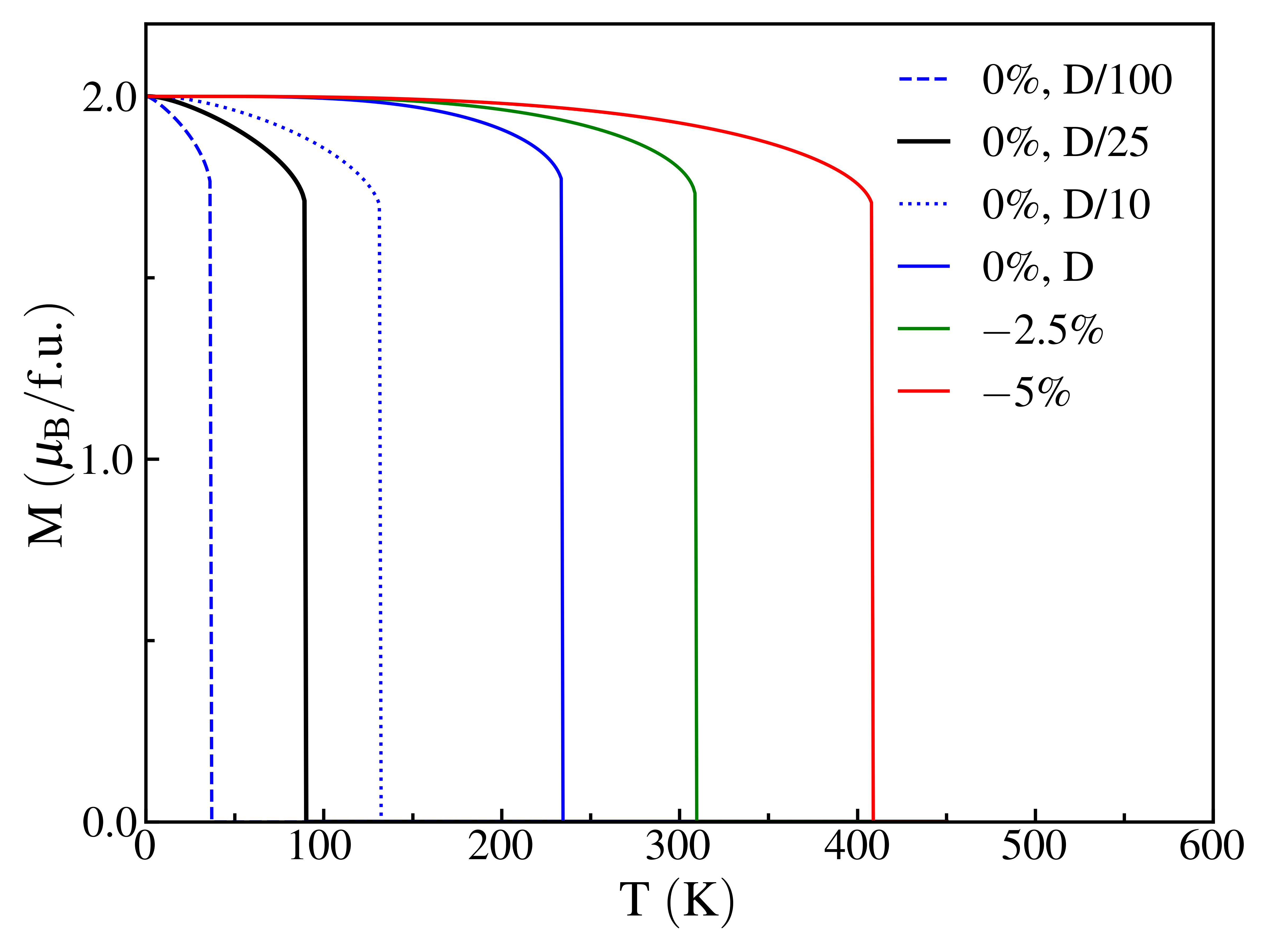}
  \caption{The magnetization $M$ as a function of temperature calculated by RSWT. Here the SIA parameter is $D$=25 meV. $T_ {\rm C}$ is determined by $M$ = 0.}
  \label{RSWT}
\end{figure}

Using the three FM exchange parameters and the huge SIA parameter $D$ = 25 meV, here we perform RSWT calculations (see Section VI in SM~\cite{SM}) and find that FeS$_2$ monolayer has a pretty high $T_{\rm C}$ = 261 K as seen in Fig. \ref{RSWT}. In the RSWT calculations, the $T_{\rm C}$ is sensitive to the SIA energy, and $T_{\rm C}$ would drop a lot when the SIA parameter is reduced from the present huge value of 25 meV down to the more common few meV or tenths of meV: e.g., $T_{\rm C}$ is reduced to 132 K for SIA=2.5 meV, and down to 37 K for SIA=0.25 meV (see Fig. \ref{RSWT}). These results show that the present huge SIA indeed contributes a lot to stabilize the pretty high $T_{\rm C}$ = 261 K. Note that when using the three FM parameters ($J_1$ = 4.73 meV, $J_2$ = 1.33 meV, and $J_3$ = 1.27 meV, see Table S7 in SM) and the unchanged $D$ = 25 meV both obtained by LSDA+SOC+$U$ with $U$ = 6 eV, the $T_{\rm C}$ is somewhat reduced to 234 K. When $U$ = 4 eV is assumed, the small gap is closed, and the FM metallic solution gives a stronger itinerant FM behavior and higher $T_{\rm C}$ $>$ 261 K (compared with $U$ = 5 eV). Therefore, for the very reasonable $U$ = 5 eV (or somewhat larger $U$) to open the experimental small gap of FeS$_2$ monolayer, the $T_{\rm C}$ is 261 K (or somewhat lower but it is still a pretty high $T_{\rm C}$).

As the SIA arises from the LS $L_{z+}$ ground state which benefits from the half-filled lower $e_g^{\pi}$ doublet than the $a_{1g}$ singlet in the global trigonal crystal field, one would expect that a biaxial compressive strain would force the FeS$_6$ octahedra to elongate along the global $z$-axis, i.e., along the local cubic [111] direction, and then enlarge the crystal-field splitting between the lower $e_g^{\pi}$ doublet and the higher $a_{1g}$ singlet. As a result, the LS $L_{z+}$ ground state, out of the lower $e_g^{\pi}$ doublet, becomes even more stable, and the in-plane FM exchange could become stronger due to the shortened bond lengths. Upon the biaxial compressive strain, the energy separation between the $L_{z+}$ ground state and the higher $a_{1g}$ singlet is indeed increased as expected (see Fig. S9 in SM~\cite{SM}), but under the tensile strain, the energy separation is reduced and even changes its sign when the tensile strain is larger than 2.5\%. Moreover, under the compressive strain, the FM ground state gets more stable than other AF states, indicating a stronger FM exchange, see Section VII in SM for more details~\cite{SM}. Then, using the enhanced FM exchanges and the huge SIA under the compressive strain, our RSWT calculations show that the $T_{\rm C}$ is increased from 261 K for the pristine FeS$_2$ monolayer to 310 K for the --2.5\% strain, and to 409 K for the --5.0\% strain, see Fig. \ref{RSWT}. We also carry out MC simulations and find the $T_{\rm C}$ is comparable to the RSWT prediction.
The corresponding $T_{\rm C}$ values are 220 K, 275 K, and 350 K according to our MC simulations, see Section VIII in SM~\cite{SM}. Both sets of results arrive at the same conclusion that the pristine FeS$_2$ monolayer would have a pretty high $T_{\rm C}$ which is even above room temperature under a few percent compressive strain. Therefore, the emerging FeS$_2$ monolayer would be an appealing high-$T_{\rm C}$ 2D Ising ferromagnet.

\subsection*{D. Sulfur Vacancy}

Now we comment on why the recent experimental $T_{\rm C}$ is only about 15 K~\cite{Zhou2022}. It may well be due to the S vacancies. Such ligand vacancies are quite common in the high valent materials~\cite{goto_2021,wexler_2021,Lu_2020} and the 2D materials~\cite{wang_2013,wang_2018,Kong_2019,Ni_2022}. 
We have now performed DFT calculations using a 3$\times$3 supercell with a single sulfur vacancy in our simulations (with the vacancy ratio of 1/18), see Fig. S12(a) in Section IX of SM~\cite{SM}. The supercell structure is relaxed using LSDA, and we compare the total energies calculated by LSDA+SOC+$U$ for the spin orientation either out-of-plane or in-plane, and find that the magnetic anisotropy energy (MAE) still favors the easy out-of-plane but drops drastically down to only 1.02 meV/Fe on average. This drastic reduction of MAE is mainly due to a lifting of the orbital degeneracy by the lattice distortion associated with the sulfur vacancy, and the further crystal field splitting largely suppresses the orbital moment and the MAE. Note that the sulfur vacancy acts like an electron donor, and the minor gap in the otherwise vacancy free prototype material closes now, see Fig. S12(b) in SM. Moreover, owing to the sulfur vacancy and the lattice distortion, there exist many inequivalent Fe sites in the supercell, and there are even much more different magnetic exchange parameters. Here, we just make a crude estimate because of this complexity. For example, for the Fe site farthest away from the S vacancy, our LSDA+SOC+$U$ calculations give the averaged first-nearest neighboring exchange parameter of 6.00 meV. This small increase of the FM coupling strength, compared with the above homogeneous $J_1$ = 5.81 meV, may well be due to the enhanced itineracy associated with the electron donation of the sulfur vacancy. However, for the Fe site closest to the S vacancy, our LSDA+SOC+$U$ calculations give the averaged first-nearest neighboring exchange parameter to be --5.70 meV, which even turns into an AF type due to the S vacancy. Therefore, the averaged exchange (most likely FM) among the many different magnetic channels in the supercell with the sulfur vacancy can be expected to be not stronger than those homogeneous ones in the ideal lattice. Taking the above $J_1$, $J_2$, and $J_3$ as the upper limit, solely the drastic decrease of the MAE already significantly reduces the $T_{\rm C}$ down to 90 K (see the solid black curve in Fig. 5). Such $T_{\rm C}$ would be further reduced (and approach the experimental 15 K) when the averagely decreasing FM exchange (and even AF type for some exchange channels) is used.
Therefore, to achieve the ideal high $T_{\rm C}$ in FeS$_2$ monolayer, the S vacancy issue should be avoided during the sample preparation/growth, e.g., using the high pressure~\cite{Yan_2022},  post-growth treatment~\cite{zhao_2020} and sulfur-rich growth~\cite{zheng_2020}.

\section*{IV. Summary}

In summary, we propose FeS$_2$ monolayer to be an appealing 2D high-$T_{\rm C}$ Ising FM magnet, using density functional calculations, crystal field level analyses and spin-orbital-state diagrams, RSWT and MC simulations.
Our results reveal that the unusual high valent Fe$^{4+}$ is in the LS $L_{z+}$ spin-orbital ground state, resulting in a large orbital moment of about 1 $\mu_{\rm B}$ and huge MA of 25 meV/Fe. The negative charge transfer character associated with the Fe high valence, strong Fe 3$d$-S 3$p$ hybridization, wide bands but a small band gap all help to establish a strong FM superexchange.
As the compressive strains can further stabilize the $L_{z+}$ ground state and enhance the FM couplings, our RSWT simulations show that the $T_{\rm C}$ of FeS$_2$ monolayer is increased from 261 K for a bare monolayer up to 310$\sim$409 K under --2.5$\sim $--5\% strains. This work highlights the exploration of the spin-orbital degrees of freedom to produce the strong Ising magnetism and FM coupling. This approach may pave the way for discovering more 2D high-$T_{\rm C}$ FM materials suitable for spintronic applications.

\section*{Acknowledgements}
This work was supported by National Natural Science Foundation of China (Grants No. 12104307, No. 12174062, and No.12241402). K. Yang and Y. Ma contributed equally to this work.




\bibliography{FeS2}

\newpage
\begin{appendix}
\setcounter{figure}{0}
\setcounter{table}{0}
\renewcommand{\thefigure}{S\arabic{figure}}
\renewcommand{\thetable}{S\arabic{table}}
\renewcommand{\theequation}{S\arabic{equation}}
\renewcommand{\tablename}{Table}
\renewcommand{\figurename}{Fig.}

\titleformat*{\section}{\normalfont\Large\bfseries}
\section*{Supplemental Material for "FeS$_2$ monolayer: a high valence and high-$T_{\rm C}$ Ising ferromagnet"}

  \subsection*{\textbf{I. The trigonal crystal field and orbital state in FeS$_2$ monolayer}}

  \begin{figure}[H]
    \centering
   \includegraphics[width=8cm]{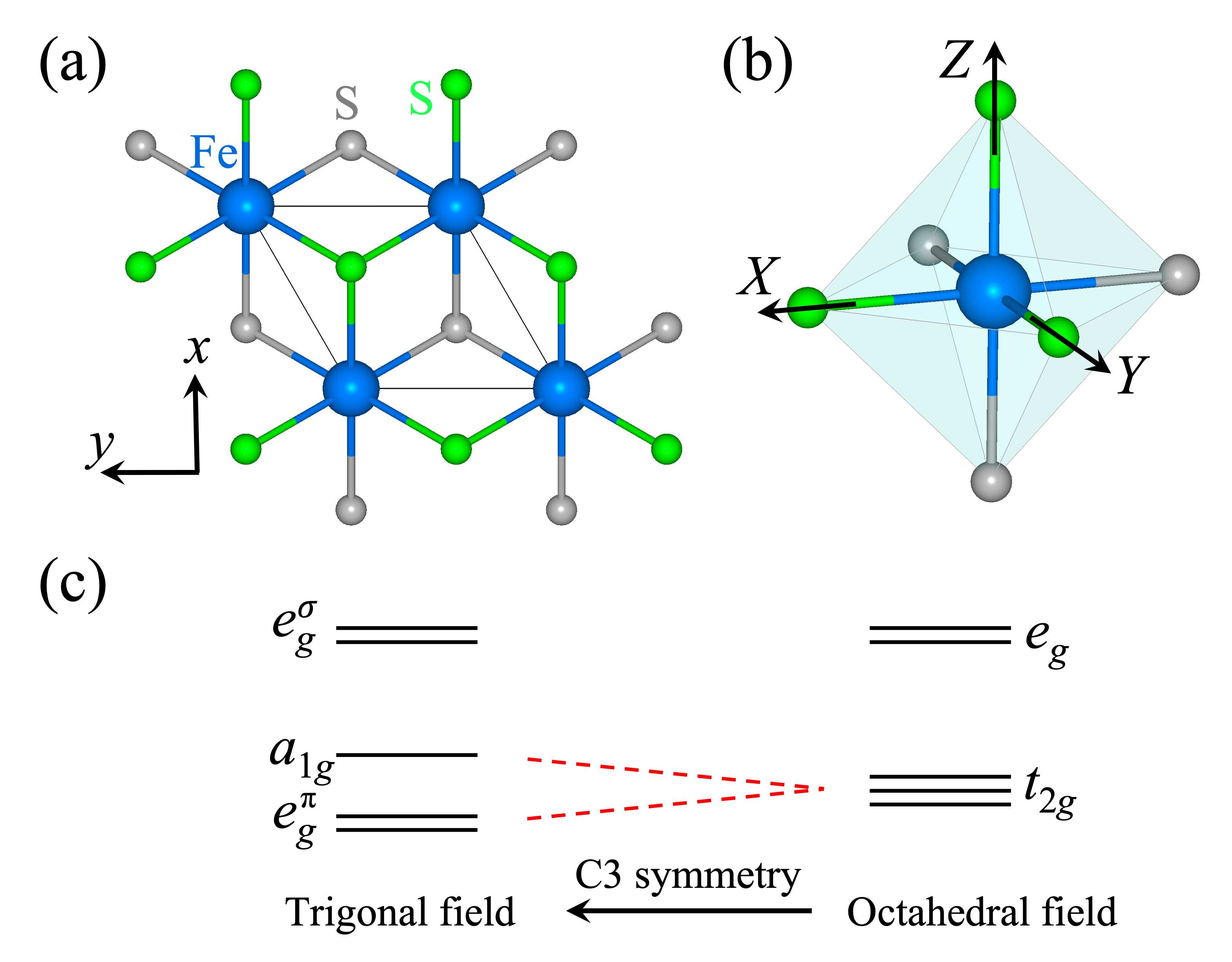}
   \centering
   \caption{(a) The crystal structure of the newly synthesized monolayer phase FeS$_2$ in the global $xyz$ coordinate system. (b) The local FeS$_6$ octahedron in its local $XYZ$ coordination system.
   (c) The Fe 3$d$ orbital states in the local octahedral but global trigonal crystal field.}
   \label{Structure}
   \end{figure}

  The crystal structure of the newly synthesized FeS$_2$ monolayer in the $P\bar{3}m1$ space group~\cite{Zhou2022} features local octahedral coordination of Fe ions and a trigonal crystal field in the global coordinate system. This crystal field splits the degenerate $t_{2g}$ triplet into the $a_{1g}$ singlet and $e_g^{\pi}$ doublet. In the following calculations, we use the global coordinate system with the $z$ axis along the [111] direction of the local FeS$_6$ octahedra and the $y$ axis along the [1$\overline{1}$0] direction (the $x$-axis is uniquely defined and the $xy$ plane is in the hexagonal $ab$ plane).
  Then the Fe 3$d$ wave functions under the global $xyz$ coordinate system in trigonal crystal field can be written as
  \begin{equation}
  \begin{aligned}
  &\left|a_{1 g}\right\rangle=\left|3 z^{2}-r^2\right\rangle\\
  &\left|e_{g_1}^{\pi}\right\rangle=\sqrt{\frac{2}{3}}\left|x^{2}-y^{2}\right\rangle-\sqrt{\frac{1}{3}}\left|xz\right\rangle \\
  &\left|e_{g_2}^{\pi}\right\rangle=\sqrt{\frac{1}{3}}\left| yz\right\rangle+\sqrt{\frac{2}{3}}\left|xy\right\rangle \\
  &\left|e_{g_1}^\sigma\right\rangle=\sqrt{\frac{1}{3}}\left|x^{2}-y^{2}\right\rangle+\sqrt{\frac{2}{3}}\left|xz\right\rangle \\
  &\left|e_{g_2}^\sigma\right\rangle=\sqrt{\frac{2}{3}}\left|yz\right\rangle-\sqrt{\frac{1}{3}}\left|xy\right\rangle
  \end{aligned}
  \label{eq1}
  \end{equation}
  When the SOC is considered, different linear combinations of the $a_{1g}$ singlet and $e_g^{\pi}$ doublet yield the $L$ =1 orbital moment:
  
  \begin{equation}
    \begin{aligned}
    &\left|L_{z\pm}\right\rangle=\frac{1}{\sqrt{2}}\left(\left|e_{g 1}^\pi\right\rangle\pm i\left|e_{g_2}^\pi\right\rangle\right) \\
    &\left|L_{x\pm}\right\rangle=\frac{1}{\sqrt{2}}\left(\left|e_{g_2}^\pi\right\rangle\pm i\left|a_{1 g}\right\rangle\right) \\
    &\left|L_{y\pm}\right\rangle=\frac{1}{\sqrt{2}}\left(\left|e_{g_1}^\pi\right\rangle\pm i\left|a_{1 g}\right\rangle\right)
    \end{aligned}
  \label{eq2}
    \end{equation}
  where $L_{x, y, z}$ represents the orientation of the orbital moment along the $x$, $y$, and $z$ axes, and $\pm$ stands for the $L_i$ = $\pm$1. Owing to the degeneracy of the $e_g^{\pi}$ doublet in the trigonal crystal field, the $L_x$ and $L_y$ states are still degenerate.

  \subsection*{\textbf{II. The LSDA+SOC+$U$ DOS results for the different spin-orbtial states}}

  \renewcommand\arraystretch{1.5}
  \begin{table}[H]
    \centering
  \caption{Relative total energies $\Delta$\textit{E} (meV/fu), local spin and orbital moments ($\mu_{\rm B}$) in different spin-orbital states calculated by LSDA+SOC+$U$ with $U$ = 5 eV. The symbols $\perp$ ($\parallel$) stand for the out-of-plane (in-plane) magnetization.
  }
  \begin{tabular}{l@{\hskip5mm}r@{\hskip5mm}c@{\hskip5mm}r}
  \hline\hline
  States & $\Delta$\textit{E} & Fe$_{\rm spin}$  & Fe$_{\rm orb}$   \\ \hline
  $L_{z+}$ $\perp$     &   0   &   1.98    &   0.94    \\
  $L_{z+}$ $\parallel$ &  25   &   1.98    &   0.27    \\
  $L_{x+}$ $\parallel$ &  42   &   2.00    &   1.00    \\
  $a_{1g}$ $\parallel$ &	44   &   1.88	 &   0.28    \\
  $L_{z-}$ $\perp$     &  61   &	 1.99	 & --0.84    \\
  \hline\hline
   \end{tabular}
   \label{tb1}
  \end{table}
  \renewcommand\arraystretch{1.5}
  \begin{table}[H]
    \centering
  \caption{Relative total energies $\Delta$\textit{E} (meV/fu), local spin and orbital moments ($\mu_{\rm B}$) in different spin-orbital states calculated by LSDA+SOC+$U$  with $U$ = 6 eV. The symbols $\perp$ ($\parallel$) stand for the out-of-plane (in-plane) magnetization.
  }
  \begin{tabular}{l@{\hskip5mm}r@{\hskip5mm}c@{\hskip5mm}r}
  \hline\hline
  States & $\Delta$\textit{E} & Fe$_{\rm spin}$  & Fe$_{\rm orb}$   \\ \hline
  $L_{z+}$ $\perp$     &   0   &   2.10    &   1.03    \\
  $L_{z+}$ $\parallel$ &  25   &   2.12    &   0.38    \\
  \hline\hline
   \end{tabular}
   \label{tb1}
  \end{table}

  \begin{figure}[H]
    \centering
  \includegraphics[width=8cm]{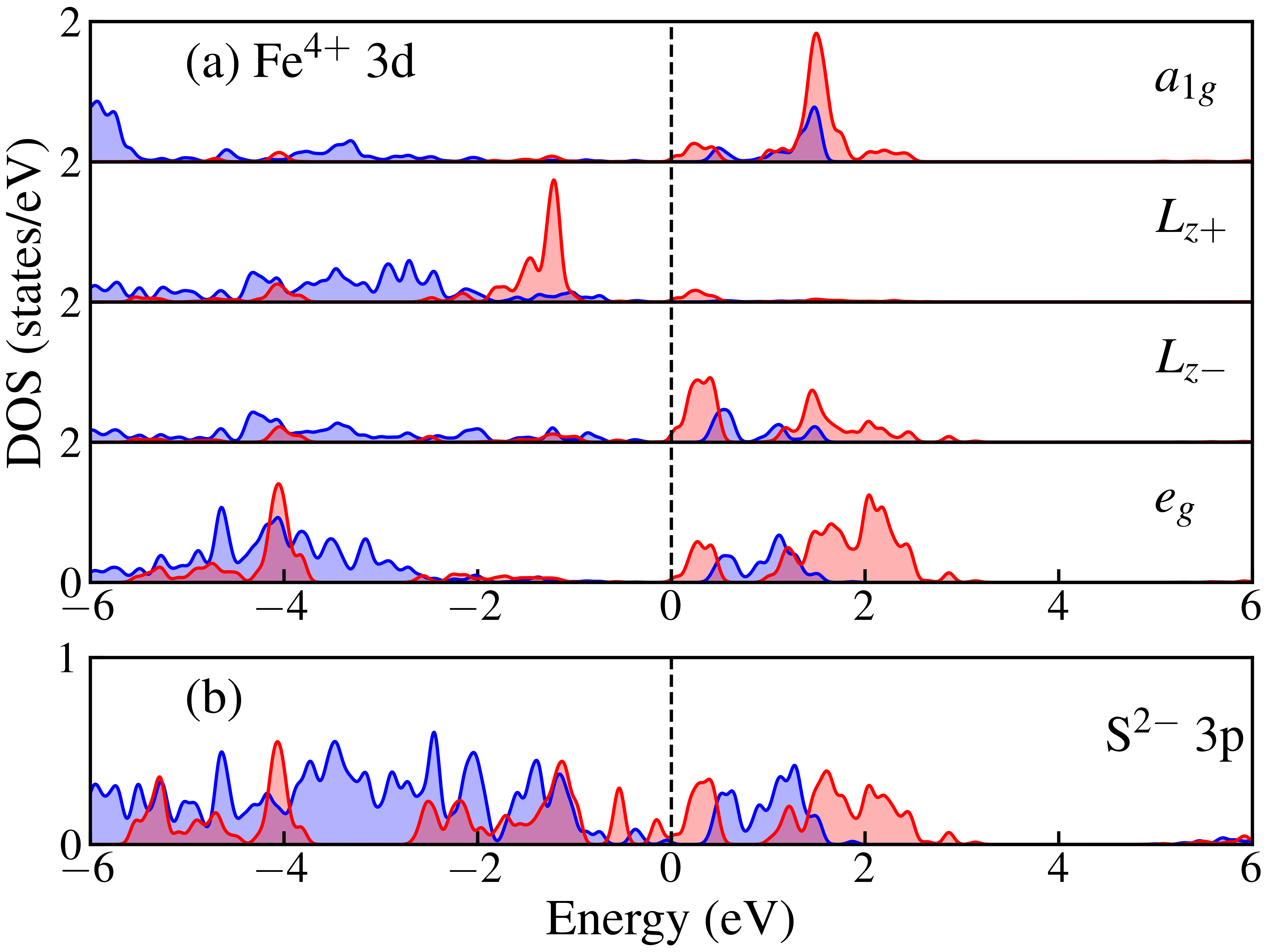}
  \centering
   \caption{(a) Fe 3$d$ and (b) S 3$p$ DOS for $L_{z+} \parallel$ state. The blue (red) curves stand for the up (down) spin channel.
  }
  \label{states}
  \end{figure}
  
  \begin{figure}[H]
    \centering
  \includegraphics[width=8cm]{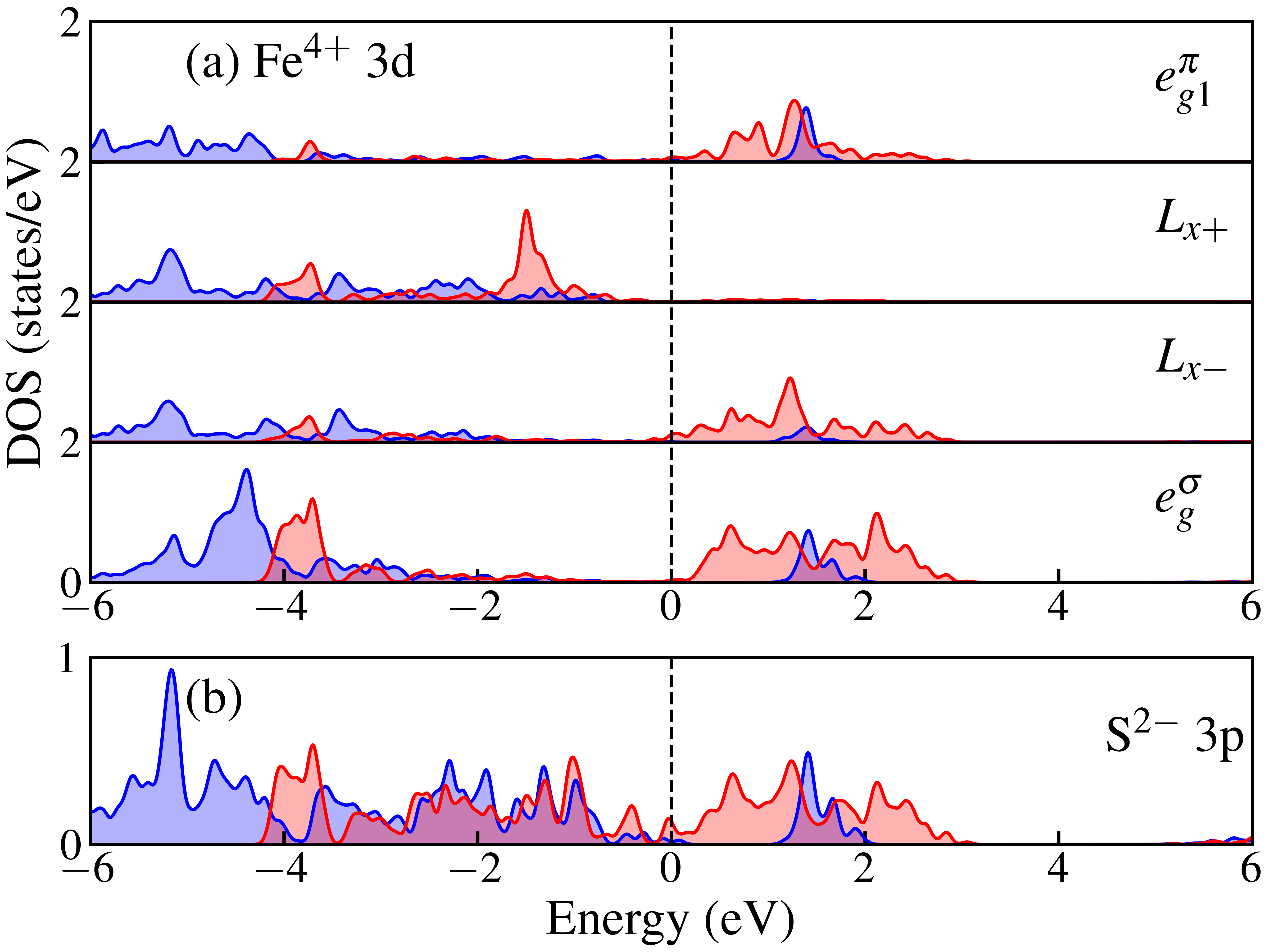}
  \centering
   \caption{(a) Fe 3$d$ and (b) S 3$p$ DOS for $L_{x+}\parallel$ state.
  }
  \label{states}
  \end{figure}
  
  \begin{figure}[H]
    \centering
  \includegraphics[width=8cm]{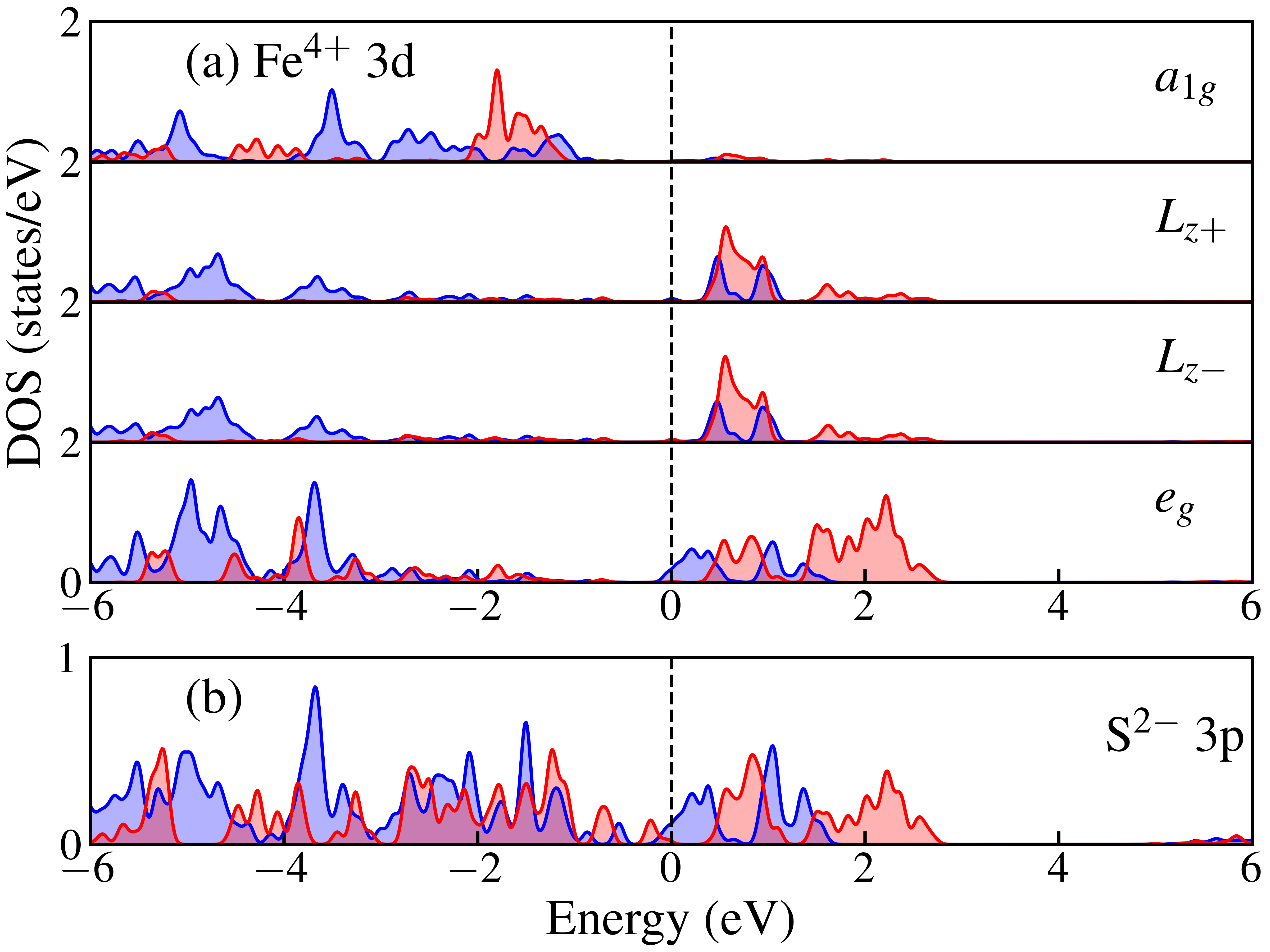}
  \centering
   \caption{(a) Fe 3$d$ and (b) S 3$p$ DOS for $a_{1g}\parallel$ state.
  }
  \label{states}
  \end{figure}
  \begin{figure}[H]
    \centering
  \includegraphics[width=9cm]{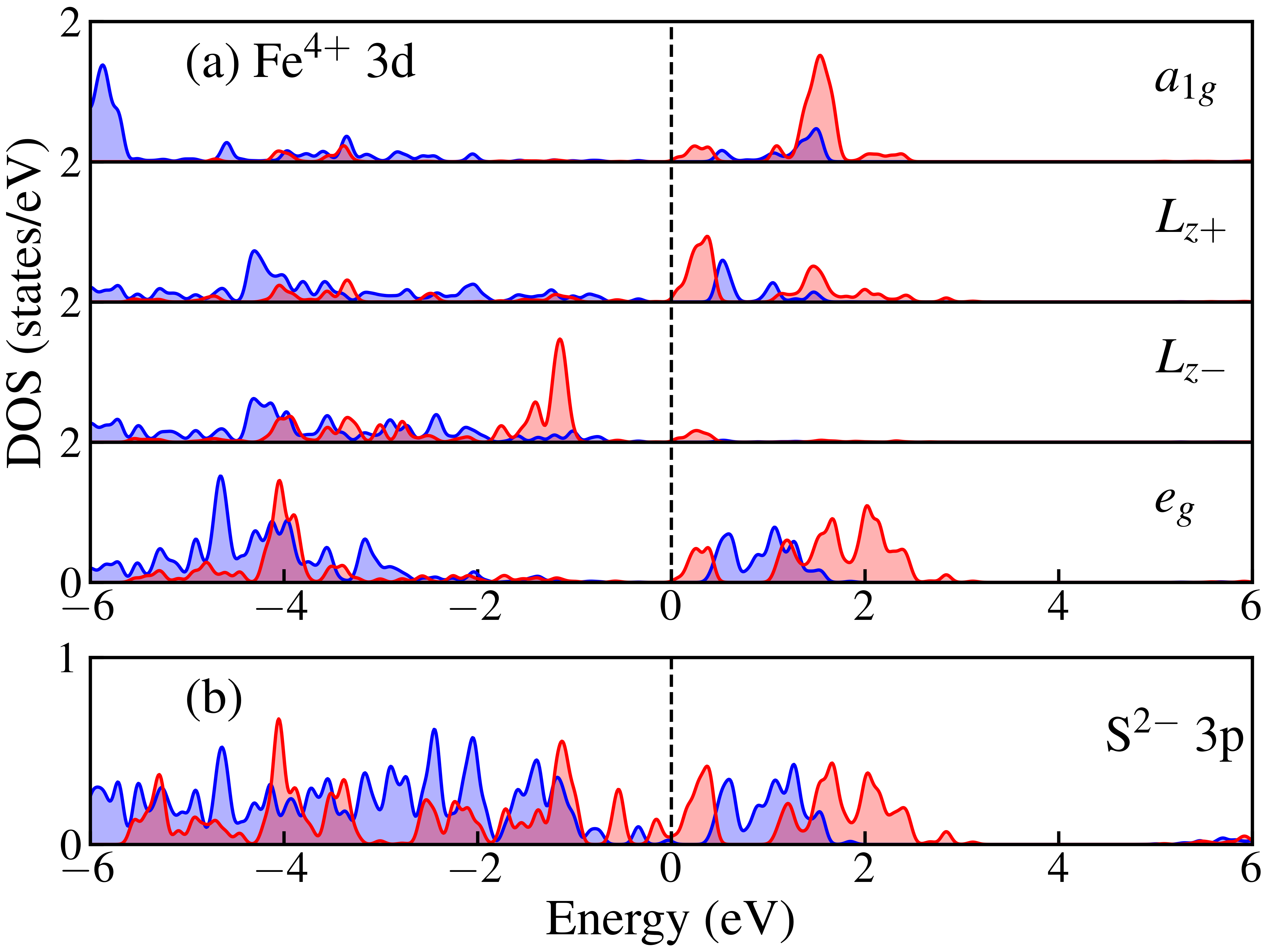}
  \centering
   \caption{(a) Fe 3$d$ and (b) S 3$p$ DOS for $L_{z-}$ state.
  }
  \label{states}
  \end{figure}

  \subsection*{\textbf{III. The hybrid functional band calculations}}
  
  The hybrid functional band structure for the FeS$_2$ monolayer is calculated using the full-potential augmented plane wave plus local orbital code (WIEN2K)\cite{WIEN2k}. The muffin-tin sphere radii are chosen to be 2.2 and 2.0 bohrs for Fe, and S atoms, respectively. The cutoff energy of 12 Ry is used for plane wave expansion. To account for the electron correlation effect of Fe 3$d$ electrons, we employ a hybrid functional with a quarter Hartree-Fock exchange mixed into LSDA\cite{Becke_1993,Becke_1996, Fock25, hf_1,hf_2}. SOC is included by the second variational method with scalar relativistic wave functions, and the magnetization direction is set along the $c$-axis.
  
  \begin{figure}[H]
    \centering
  \includegraphics[width=9cm]{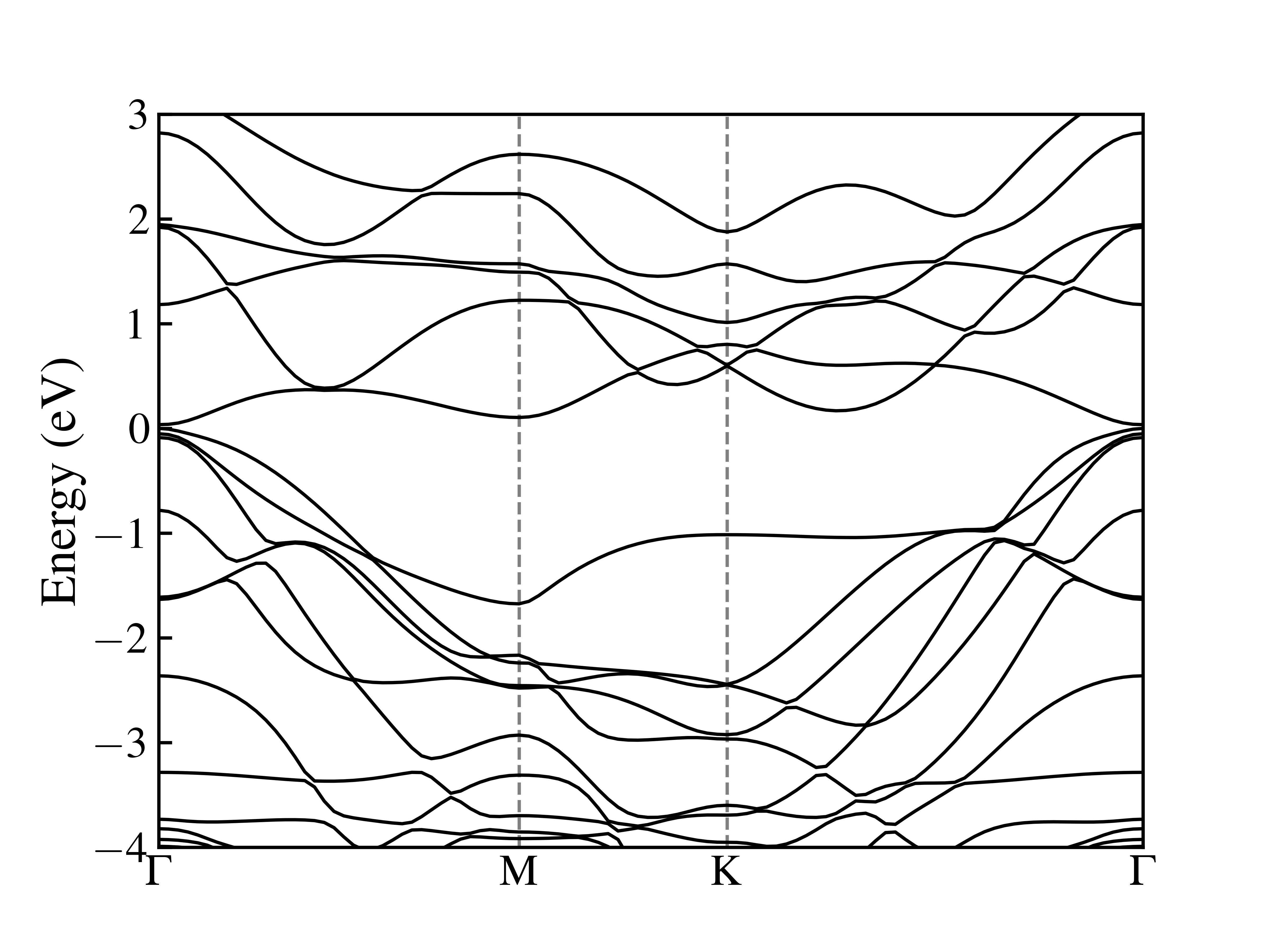}
  \centering
   \caption{The calculated band structure with a tiny band gap of 36 meV by hybrid functional calculation.
  }
  \end{figure}

  \subsection*{\textbf{IV. The exchange parameters}}

    \begin{figure}[H]
      \includegraphics[width=9cm]{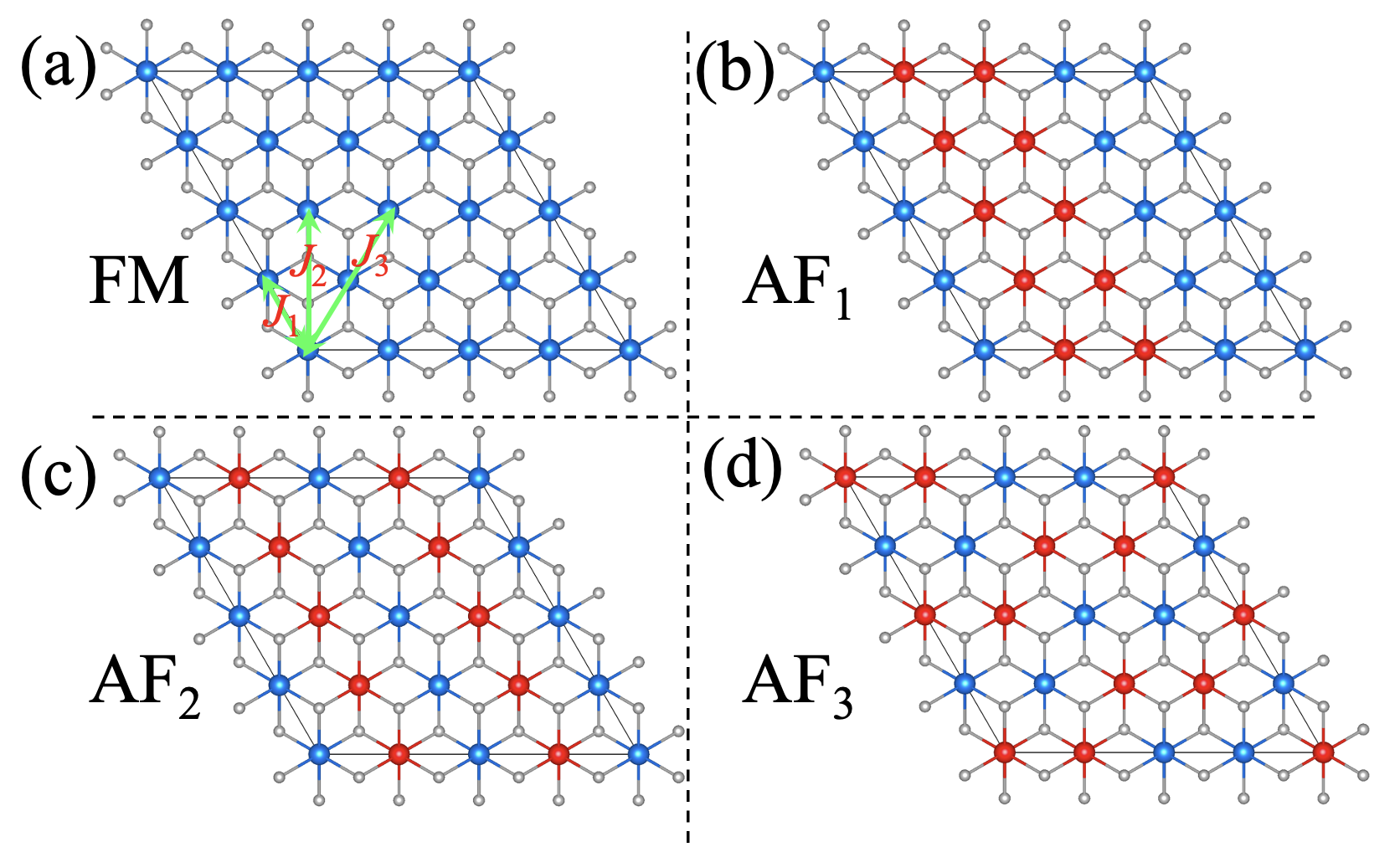}
      \centering
       \caption{The four common commensurate magnetic structures of FeS$_2$ monolayer. The blue (red) balls stand for the up (down) spin Fe ions.
      }
      \label{spin_states}
      \end{figure}

  In order to study the magnetic properties of the FeS$_2$ monolayer, a 4$\times$4 supercell is selected within the LSDA+SOC+$U$ framework. The four common magnetic structures are used to estimate the three exchange parameters: $J_1$, $J_2$, and $J_3$,
  see Fig \ref{spin_states}(a).
  The results show that the FM solution is the most stable and has a lower total energy than the three antiferromagnetic (AF) states by 24$\sim $32 meV/f.u., see Table \ref{tb2}.
  Counting $-JS^2$ for each pair of LS Fe$^{4+}$ $S$ = 1 ions (positive $J$ refers to FM exchange), the magnetic exchange energies of the FM state and the three AF states per formula unit are written as follows:
  
    \begin{equation}
      \begin{array}{ll}
        E_{\text{FM}} &= (-3J_1-3J_2-3J_3)S^2\\
        E_{\text{AF}_1} &= (-J_1+J_2+J_3)S^2\\
        E_{\text{AF}_2} &= (+J_1+J_2-3J_3)S^2\\
        E_{\text{AF}_3} &= (+J_1-J_2+J_3)S^2
      \end{array}
      \label{eq3}
    \end{equation}

  \renewcommand\arraystretch{1.3}
    \begin{table}[H]
    \centering
    \caption{Relative total energies $\Delta$\textit{E} (meV/fu), local spin and orbital moments ($\mu_{\rm B}$) for the FeS$_2$ monolayer by LSDA+$U$+SOC with $U$ = 5 eV. The derived three exchange parameters $J_1$ = 5.81 meV, $J_2$ = 2.01 meV, and $J_3$ = 1.05 meV are all FM.
    }
  
    \begin{tabular}{c@{\hskip5mm}c@{\hskip5mm}r@{\hskip5mm}r@{\hskip5mm}}
    \hline\hline
    States & $\Delta$\textit{E} & Fe$_{\rm spin}$  & Fe$_{\rm orb}$    \\ \hline
    FM      &  0              &  1.97       &  0.94          \\
    AF$_1$    &  24.23        &  $\pm$1.95  &  $\pm$1.05      \\
    AF$_2$   &  31.64         &  $\pm$1.91  &  $\pm$1.09     \\
    AF$_3$    &  31.65        &  $\pm$1.95  &  $\pm$1.03      \\
    \hline\hline
     \end{tabular}
     \label{tb2}
    \end{table}

    \renewcommand\arraystretch{1.3}
    \begin{table}[H]
    \centering
    \caption{Relative total energies $\Delta$\textit{E} (meV/fu), local spin and orbital moments ($\mu_{\rm B}$) for the FeS$_2$ monolayer by LSDA+$U$+SOC with $U$ = 6 eV. The derived three exchange parameters $J_1$ = 4.73 meV, $J_2$ = 1.33 meV, and $J_3$ = 1.27 meV are all FM.
    }
  
    \begin{tabular}{c@{\hskip5mm}c@{\hskip5mm}r@{\hskip5mm}r@{\hskip5mm}}
    \hline\hline
    States & $\Delta$\textit{E} & Fe$_{\rm spin}$  & Fe$_{\rm orb}$    \\ \hline
    FM      &  0              &  2.06       &  1.03          \\
    AF$_1$    &  19.91       &  $\pm$2.10  &  $\pm$1.12      \\
    AF$_2$   &  24.30         &  $\pm$2.03  &  $\pm$1.11     \\
    AF$_3$    &  26.71        &  $\pm$2.08  &  $\pm$1.15      \\
    \hline\hline
     \end{tabular}
     \label{tb2}
    \end{table}

  \subsection*{\textbf{V. Two major superexchange channels}}
  
\begin{figure}[H]
    \includegraphics[width=9cm]{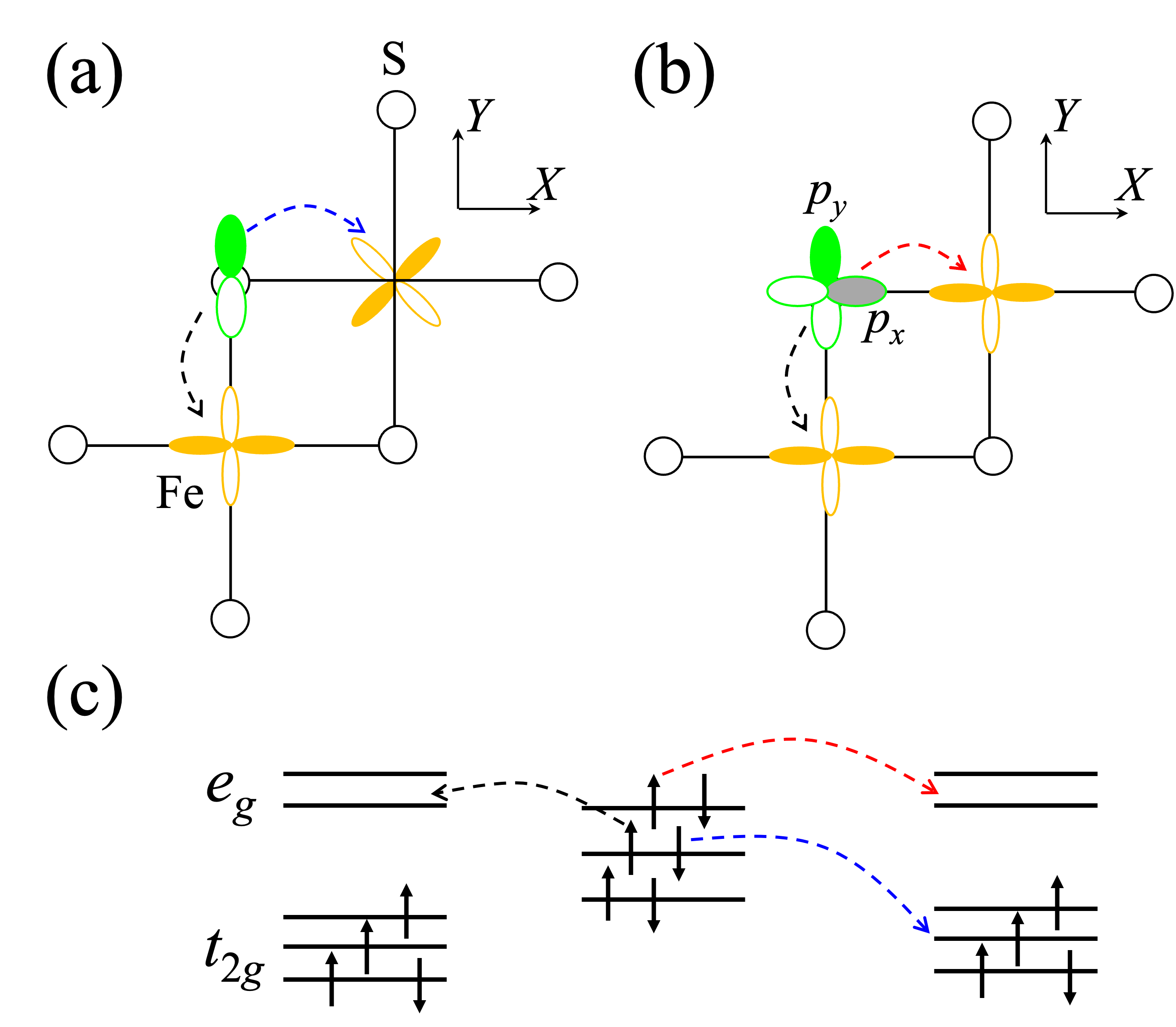}
    \centering
     \caption{Schematic plot of the (near-) 90$^\circ$  FM superexchange channels: (a) ($XY$)-$p_Y$-($X^2-Y^2$), and (b)  ($X^2-Y^2$)-($p_X$, $p_Y$)-($X^2-Y^2$).
    }
    \label{supE}
    \end{figure}

  The two major superexchange channels, counting the large $pd\sigma$ and medium $pd\pi$ hybridizations, contribute to the FM $J_1$ coupling.
  One arises from $pd\pi$ hybridization via the same $p_Y$ orbital (but in an opposite spin orientation, figs. \ref{supE}(a) and (c)), and the other involves the $pd\sigma$ hybridization via the orthogonal ($p_X$, $p_Y$)orbitals (Figs. \ref{supE}(b) and (c)).

  \subsection*{\textbf{VI. The  renormalized spin-wave theory (RSWT) to simulate the $T_{\rm C}$}}
  
  We determine the $T_{\rm C}$ of $FeS_2$ monolayer with renormalized spin-wave theory\cite{li_2018}.
  One of the key advantages of using the renormalized spin-wave theory (RSWT) to simulate the critical temperature ($T_{\rm C}$) of a magnetic material is its ability to take into account the effects of quantum and thermal fluctuations on the spin waves. These effects are often neglected in the standard spin-wave theory, but they can have a significant impact on the behavior of magnetic materials at low temperatures.
  By accurately accounting for these effects, RSWT can provide more accurate predictions of $T_{\rm C}$ for a wide range of magnetic materials, including itinerant and localized systems, as well as materials with different types of spin interactions. This can be useful for understanding the magnetic properties of these materials and for designing new magnetic materials with desired properties.
  Additionally, RSWT is a relatively simple and computationally efficient approach, which makes it suitable for analyzing large systems and for performing parametric studies. This can be useful for exploring the effects of different factors on the critical temperature, such as the strength of the spin interactions or the presence of disorder in the material.
  
  The spin operator $\mathbf{S}_{lv}$ on site $v$ of $l$th unit cell can be mapped into creation and annihilation operators of magnon by Holstein-Primakoff transformation. In the linear spin-wave theory(LSWT), which is the linear approximation of the Hamiltonian, the spin operators are mapped by linear term of Holstein-Primakoff transformation, i.e. $S_{l v}^{+} \approx \sqrt{2 S} a_{l v}$, $S_{l v}^{-} \approx \sqrt{2 S} a_{l v}^{\dagger}$ and $S_{l v}^z=S-a_{l v}^{\dagger} a_{l v}$. Thus, the Hamiltonian, which only contains one-body operators, can be solved by Fourier transformation and matrix diagonalization. In the basis of LSWT, RSWT contains the second order approximation of Hamiltonian, where spin operators are mapped by second order approximated Holstein-Primakoff transformation
  
  \begin{equation}
  \begin{array}{ll}
  S_{l v}^{+} &\approx \sqrt{2 S}\left(a_{l v}-\frac{a_{l v}^{\dagger} a_{l v} a_{l v}}{4 S}\right) \\
  S_{l v}^{-} &\approx \sqrt{2 S}\left(a_{1 v}^{\dagger}-\frac{a_{l v}^{+} a_{l v}^{\dagger} a_{l v}}{4 S}\right) \\
  S_{l v}^z&=S-a_{l v}^{\dagger} a_{l v}.
  \end{array}
  \end{equation}
  
  We define the number operator $\langle n_{kv}\rangle$, where $k$ is wave vector in Fourier transformation, as our mean-field order parameter. Under Hartree-Fork approximation, we can diagonalize the Hamiltonian and obtain the self-consistent equation, which can be solved recursively. And following the Bose-Einstein statistics, the magnetization is expressed as
  
  \begin{equation}
  \frac{M}{M_0}=1-\frac{1}{n N S} \sum_k\left\langle n_k\right\rangle.
  \end{equation}

   \subsection*{\textbf{VII. The strain results calculated by LSDA+SOC+$U$}}
  
  \begin{figure}[H]
    \includegraphics[width=8cm]{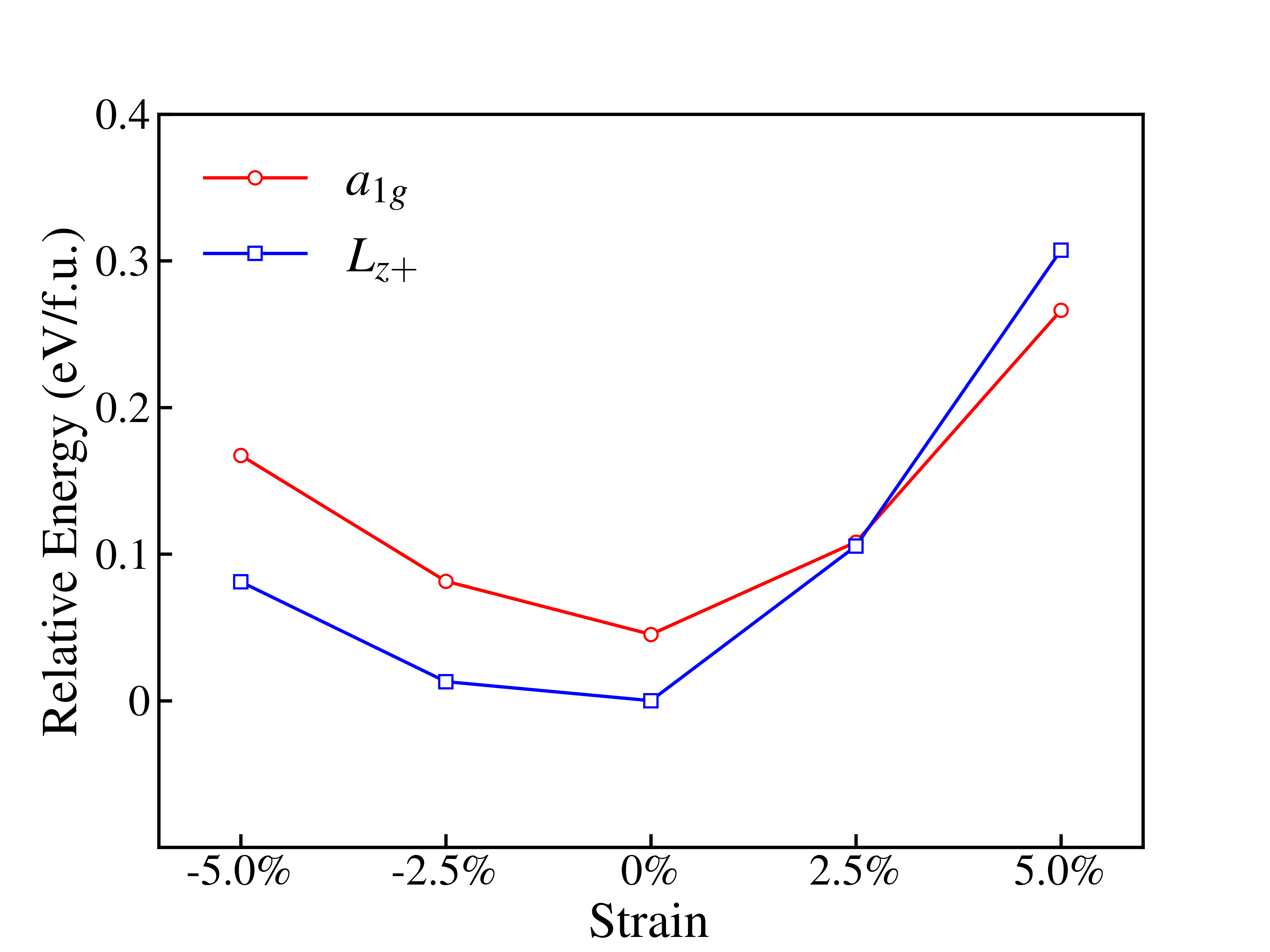}
    \centering
    \caption{The relative total energy (eV/f.u.) of the $L_{z+}$ ground state against the $a_{1g}$ state under different strains.}
    \label{strain}
  \end{figure}
  
  \renewcommand\arraystretch{1.5}
  \begin{table}[H]
    \centering
  \caption{Relative total energies $\Delta$\textit{E} (meV/fu), local spin and orbital moments ($\mu_{\rm B}$), and total spin moments ($\mu_{\rm B}$/fu) for different states of FeS$_2$ under different strain calculated by LSDA+SOC+$U$. The perpendicular magnetization is assumed in most cases, and the in-plane ($\parallel$) magnetization is also set for some cases.
  }
  \begin{tabular}{c@{\hskip3.5mm}|c@{\hskip3.5mm}c@{\hskip3.5mm}c@{\hskip3.5mm}l@{\hskip3.5mm}r@{\hskip3.5mm}c@{\hskip3.5mm}}
  \hline\hline
  Strain & States & $\Delta$\textit{E} & Fe$_{\rm spin}$  & Fe$_{\rm orb}$  & S$_{\rm spin}$ &Total$_{\rm spin}$  \\ \hline
  \multirow{3}{*}{-5\%}
   & $L_{z+}$             &  0    &   1.82    &   0.92        &   0.03   &  1.92    \\
   & $L_{z+}$ $\parallel$ &  26   &   1.76    &   0.11 ($x$)       &  0.04   &  1.87    \\
   & $a_{1g}$             &  86   &   1.87    &	0.03 &  0.02 	 &  1.97	  \\
  \hline
  \multirow{3}{*}{-2.5\%}
   & $L_{z+}$             &  0    &   1.84    &   0.95        &   0.02   &  1.91    \\
   & $L_{z+}$ $\parallel$ &  27   &   1.85    &   0.23 ($x$)       &  0.02   &  1.91    \\
   & $a_{1g}$             &  69   &   1.76	    &	-0.04 &  0.02 	 &  1.82	  \\
  \hline
  \multirow{3}{*}{0\%}
   & $L_{z+}$             &  0    &   1.98    &   0.94        &   0.00   &  2.00    \\
   & $L_{z+}$ $\parallel$ &  25   &   1.98    &   0.27 ($x$)       &  --0.01   &  1.99    \\
   & $a_{1g}$             &  45   &   1.84	 &	  0.05       &  --0.01 	 &  1.82	  \\
  \hline
  \multirow{2}{*}{2.5\%}
   & $L_{z+}$             &  0    &   2.03    &   0.87        &   -0.02   &  1.99    \\
   & $a_{1g}$             &  3   &   1.84    &	-0.02 &  -0.02 	 &  1.82	  \\
  \hline
  \multirow{2}{*}{5\%}
   & $a_{1g}$             &  0   &   1.84	    &	0.01 &  -0.03 	 &  1.81	  \\
   & $L_{z+}$             &  41    &   2.10    &   0.75        &   -0.05   &  1.98    \\
  \hline\hline
   \end{tabular}
   \label{tb3}
  \end{table}

  \renewcommand\arraystretch{1.3}
  \begin{table}[H]
  \centering
  \caption{Relative total energies $\Delta$\textit{E} (meV/fu), local spin and orbital moments ($\mu_{\rm B}$) and total spin moments ($\mu_{\rm B}$/fu) for the FeS$_2$ monolayer under compressive strain by LSDA+$U$+SOC.
  }
  
  \begin{tabular}{c@{\hskip3.5mm}|c@{\hskip3.5mm}c@{\hskip3.5mm}r@{\hskip3.5mm}r@{\hskip3.5mm}c@{\hskip3.5mm}c@{\hskip3.5mm}}
  \hline\hline
  Strain & States & $\Delta$\textit{E} & Fe$_{\rm spin}$  & Fe$_{\rm orb}$  & S$_{\rm spin}$ &Total$_{\rm spin}$  \\ \hline
  \multirow{4}{*}{0\%}
   & FM      &  0              &  1.97       &  0.94         &  $-$0.001      &  1.98   \\
   & AF$_1$    &  24.23        &  $\pm$1.95  &  $\pm$1.05    &  $\mp$0.04  &    0.00  \\
   & AF$_2$   &  31.64         &  $\pm$1.91  &  $\pm$1.09    &  $\mp$0.01  &   0.00   \\
   & AF$_3$    &  31.65        &  $\pm$1.95  &  $\pm$1.03    &  $\mp$0.01  &    0.00     \\
  \hline
  \multirow{4}{*}{-2.5\%}
   & FM      &  0              &  1.83       &  0.96         &  $-$0.017      &  1.89   \\
   & AF$_1$    &  31.91        &  $\pm$1.86  &  $\pm$1.09    &  $\mp$0.013  &    0.00  \\
   & AF$_2$   &  29.14         &  $\pm$1.81  &  $\pm$1.09    &  $\mp$0.003  &   0.00   \\
   & AF$_3$    &  41.81        &  $\pm$1.83  &  $\pm$1.10    &  $\mp$0.011  &    0.00     \\
  \hline
  \multirow{4}{*}{-5\%}
   & FM      &  0              &  1.87       &  0.91         &  $-$0.030      &  1.97   \\
   & AF$_1$    &  50.57        &  $\pm$1.79  &  $\pm$1.07    &  $\mp$0.007  &    0.00  \\
   & AF$_2$   &  48.89         &  $\pm$1.76  &  $\pm$1.11    &  $\mp$0.001  &   0.00   \\
   & AF$_3$    &  43.83        &  $\pm$1.70  &  $\pm$0.98    &  $\mp$0.001  &    0.00     \\
  \hline\hline
   \end{tabular}
   \label{tb4}
  \end{table}

  \subsection*{\textbf{VIII. The  Monte Carlo (MC) simulations}}

  We perform a Monte Carlo (MC) simulation to simulate the thermal dynamic behavior and estimate the critical temperature.
  To choose a reasonable lattice size, three kinds of lattices are constructed in the simulations: $10\times10\times1$, $15\times15\times1$ and $20\times20\times1$. In each simulation, we use 8000 Monte Carlo steps to relax the system and then 15000 Monte Carlo steps to count magnetic moments.
  During the simulation step, each spin is rotated randomly in the three-dimensional space. The spin dynamical process is studied by the classical Metropolis methods~\cite{Metropolis}.
  And in each Monte Carlo step, more than 6000 spins are flipped. These simulations show that $15\times15\times1$ lattice is large enough for the simulation.

\begin{figure}[H]
    \centering
  \includegraphics[width=9cm]{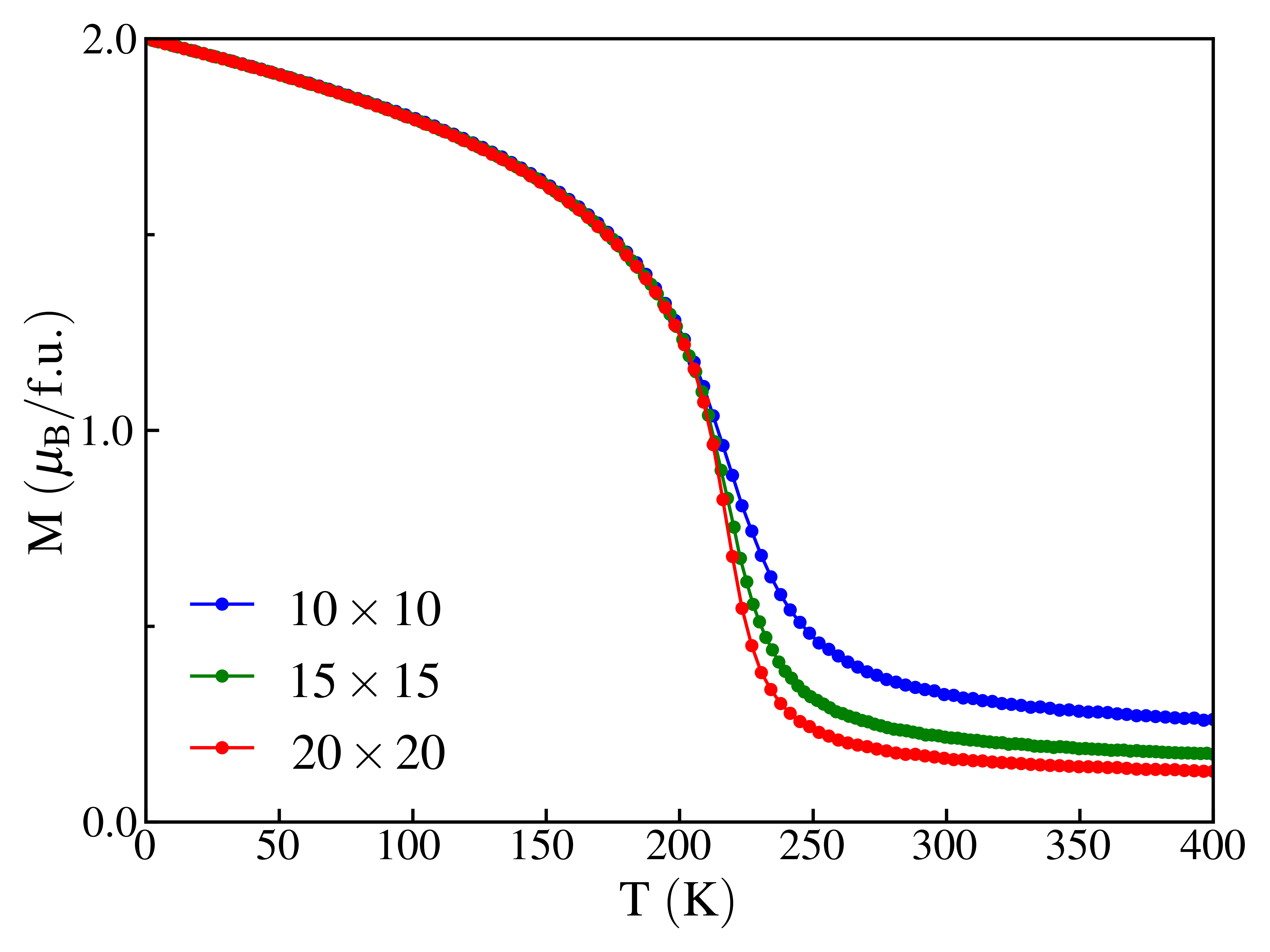}
  \centering
   \caption{Monte Carlo simulation of magnetization for FeS$_2$ monolayer in the $10\times10\times1$, $15\times15\times1$, and $20\times20\times1$ lattices.
  }
  \label{m10}
  \end{figure}

  With the above study of the FM coupling and the giant perpendicular MA,
  we now assume the following spin Hamiltonian and carry out Monte Carlo simulations
  \begin{equation}
    H=-\sum_{k = 1, 2, 3}\sum_{\langle i j\rangle} J_{k} \mathbf{S}_{i} \cdot \mathbf{S}_{j}-\sum_{i} D\left(S_{i}^{z}\right)^{2}
  \end{equation}
  where the first term describes the Heisenberg isotropic exchange (FM when $J$ $>$ 0), the second term is the SIA with the easy magnetization $z$ axis (when $D$ > 0), and $S_i$ = 1 for the LS Fe$^{4+}$ in the FeS$_2$ monolayer.
  Then, using the exchange parameters ($J_1$ = 5.81 meV, $J_2$ = 2.01 meV, and $J_3$ = 1.05 meV) and the MA energy ($D$ = 25 meV), our Monte Carlo simulations show that $T_ {\rm C}$ is 220 K for the FeS$_2$ monolayer,see Fig. \ref{m10}.
  
  \begin{figure}[H]
    \centering
  \includegraphics[width=9cm]{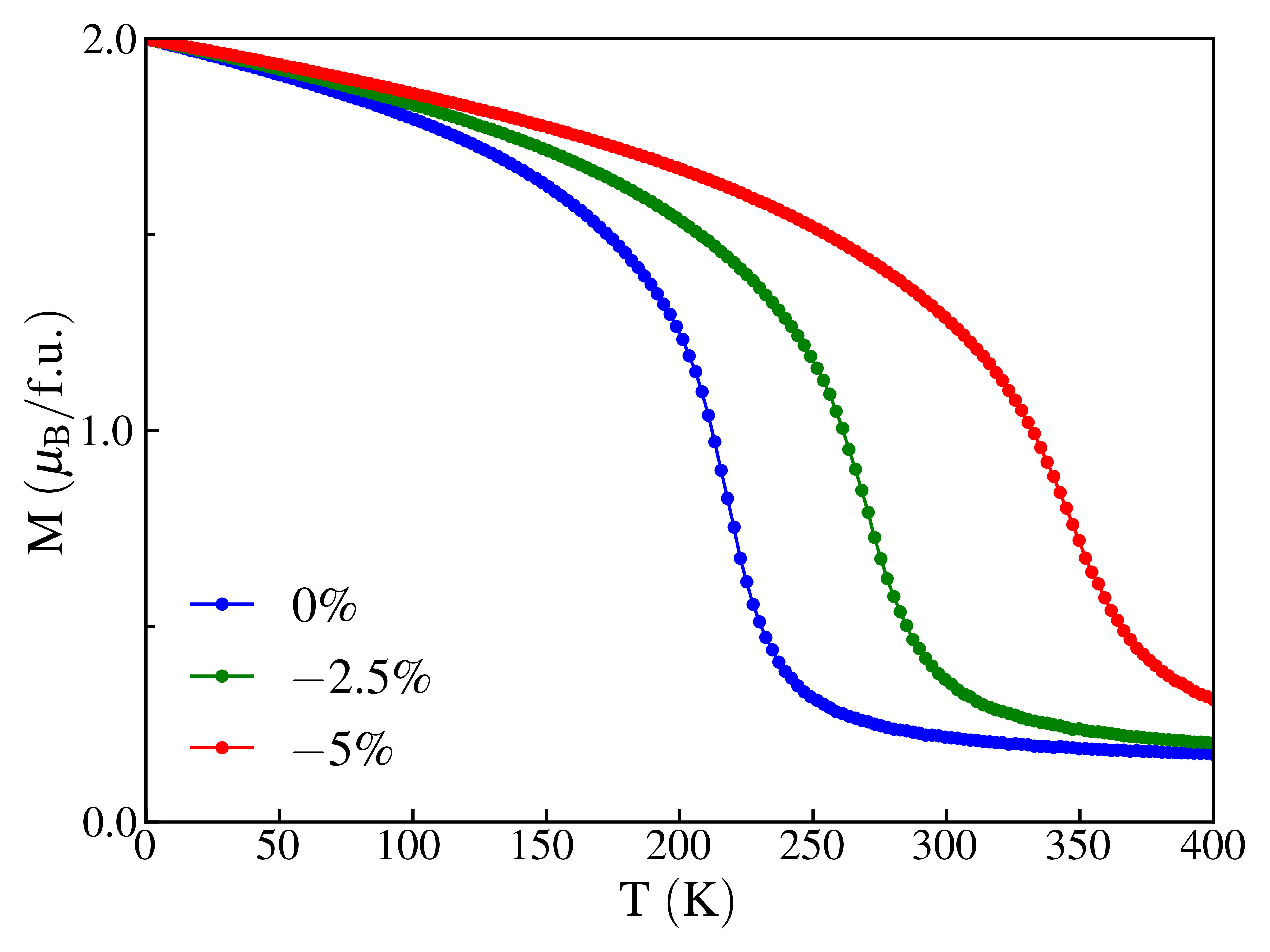}
  \centering
   \caption{Monte Carlo simulation of magnetization for FeS$_2$ monolayer in the $15\times15\times1$ lattice under different compressive strains. The $T_{\rm C}$ is increased from 220 K for a bare monolayer up to 275$\sim$350 K under --2.5$\sim $--5\% strains.
  }
  \label{m15}
  \end{figure}

  \subsection*{\textbf{IX. The sulfur vacancy calculations}}

  \begin{figure}[H]
    \centering
  \includegraphics[width=9cm]{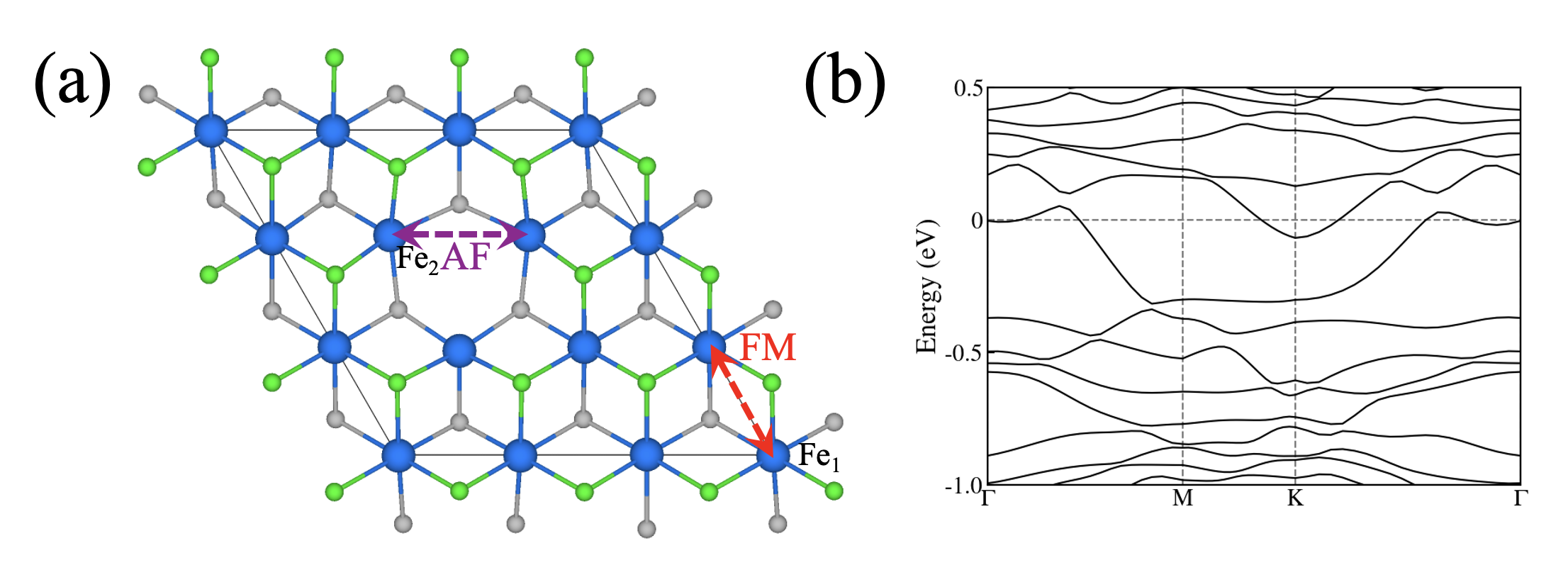}
  \centering
   \caption{(a) The 3$\times$3 supercell of FeS$_2$ with a single sulfur vacancy  (with the vacancy ratio of 1/18). Our LSDA+SOC+$U$ calculations show: for the Fe$_1$ site farthest from the vacancy, the averaged exchange parameter is 6.00 meV for the first-nearest neighboring FM couplings; for the Fe$_2$ site nearest to the vacancy, the averaged exchange parameter is –5.70 meV for the first-nearest neighboring AF couplings. (b) The corresponding band structure for the FeS$_2$ with sulfur vacancy.
  }
  \label{m15}
  \end{figure}
  

\end{appendix}

\end{document}